\documentclass[fleqn,usenatbib]{mnras}

\usepackage{newtxtext,newtxmath}
\usepackage{verbatim}

\usepackage[T1]{fontenc}
\usepackage{ae,aecompl}
\usepackage[svgnames]{xcolor}
\usepackage{hyperref}
\hypersetup{citecolor=Blue}

\usepackage{graphicx}	
\usepackage{amsmath}	
\usepackage{amssymb}	
\usepackage{amsfonts}
\usepackage{epsfig,rotating}

\def\simeq{\mathrel{\raise.3ex\hbox{$\sim$}\mkern-14mu\lower0.4ex\hbox{$-$}}}

\def\ltsima{$\; \buildrel < \over \sim \;$}
\def\simlt{\lower.5ex\hbox{\ltsima}}
\def\gtsima{$\; \buildrel > \over \sim \;$}
\def\simgt{\lower.5ex\hbox{\gtsima}}

\def\msun{{\rm M_{\odot}}}

\def\be{\begin{equation}}
\def\ee{\end{equation}}

\def\del#1{{}}
\def\SNc#1{\textcolor{red}{#1}}
\newcommand\mearth{\,{\rm M}_{\oplus}}
\newcommand\mj{\,{\rm M}_{\rm J}}
\newcommand\St{\,{\rm St}}

\title[Planet disruption in TW Hydra]{TW Hya: an old protoplanetary disc revived by its planet}

\author[Nayakshin et al]{Sergei Nayakshin$^{1} \thanks{sergei.nayakshin@le.ac.uk}$, Takashi Tsukagoshi$^2$, Cassandra Hall$^1$, Allona Vazan$^{3,4}$, 
\newauthor Ravit Helled$^4$, Jack Humphries$^1$, Farzana Meru$^{5,6}$, Patrick Neunteufel$^{1,7}$ and \newauthor Olja Panic$^8$\\
$^{1}$ School of Physics and Astronomy, University of Leicester, University Road, LE1 7RH Leicester, United Kingdom\\
$^2$ Division of Radio Astronomy, National Astronomical Observatory of Japan, Osawa 2-21-1, Mitaka, Tokyo 181-8588, Japan\\
$^3$
Racah Institute of Physics, The Hebrew University,  Jerusalem 91904, Israel\\
$^4$ Institute for Computational Science, Center for Theoretical Astrophysics \& Cosmology University of Zurich, CH-8057 Zurich, Switzerland\\
$^5$ Department of Physics, University of Warwick, Gibbet Hill Road, Coventry CV4 7AL, UK\\
$^6$ Centre for Exoplanets and Habitability, University of Warwick, Gibbet Hill Road, Coventry CV4 7AL, UK\\
$^7$ Max-Planck-Institut fur Astrophysik, Karl-Schwarzschild-Str. 1, D-85748 Garching, Germany\\
$^8$ School of Physics and Astronomy, University of Leeds, Leeds, LS2 9JT\\
}

\date{Accepted XXX. Received YYY; in original form ZZZ}

\pubyear{2019}

\begin{document}
\label{firstpage}
\pagerange{\pageref{firstpage}--\pageref{lastpage}}
\maketitle

\begin{abstract}
Dark rings with bright rims are the indirect signposts of planets embedded in protoplanetary discs.  In a recent first, an azimuthally elongated AU-scale blob, possibly a planet, was resolved with ALMA in TW Hya. The blob is at the edge of a cliff-like rollover in the dust disc rather than inside a dark ring. Here we build time-dependent models of TW Hya disc. We find that the classical paradigm cannot account for the morphology of the disc and the blob.  We propose that ALMA-discovered blob hides a Neptune mass planet losing gas and dust. We show that radial drift of mm-sized dust particles naturally explains why the blob is located on the edge of the dust disc. Dust particles leaving the planet perform a characteristic U-turn relative to it, producing an azimuthally elongated blob-like emission feature. This scenario also explains why a 10 Myr old disc is so bright in dust continuum. Two scenarios for the dust-losing planet are presented. In the first, a dusty pre-runaway gas envelope of a $\sim 40\mearth$ Core Accretion planet is disrupted, e.g., as a result of a catastrophic encounter. In the second, a massive dusty pre-collapse gas giant planet formed by Gravitational Instability is disrupted by the energy released in its massive core. Future modelling may discriminate between these scenarios and allow us to study planet formation in an entirely new way -- by analysing the flows of dust and gas recently belonging to planets, informing us about the structure of pre-disruption planetary envelopes.

\end{abstract}




\section{Introduction}

Recent advent in the high resolution imaging of protoplanetary discs via scattering light techniques and mm-continuum with ALMA  yielded many exciting examples of sub-structures in the discs, such as large scale assymetries \citep{Casassus13-banana}, spirals \citep{Benisty17-spirals}, dark and bright rings and/or gaps \citep{BroganEtal15,LongEtal18,Dsharp1,DSHARP-6}, clumps and young planets \citep{Mesa19-PDS70}. In fact it is believed that most protoplanetary discs have substructures; those that do not may simply be those that have not yet been imaged at high enough resolution \citep{Garufi18-disc-substructures}. The fact that these substructures are seen in very young discs, and that the masses of these discs appear to be insufficient to form planetary systems that we observe \citep{GreavesRice10,ManaraEtal18,WilliamsEtal19-ODISEA}, indicates that planets form very rapidly, possibly faster than 1 Myr.

On the one hand, these observations bring new challenges. The presence of mm-sized dust in a few Myr old discs is surprising because of the rapid inward radial drift of the grains \citep{Weidenschilling77,BirnstielEtal12}. While the radial drift can be slowed down by invoking very massive gas discs, $M_{\rm gas} \gtrsim (0.1-0.2)\msun$ \citep[e.g.,][]{Powell17-dust-lines,Powell19-DustLanes}, other arguments point against that. \cite{Veronesi19-light-discs} shows that the prevalence of annular rather than spiral features in many of the observed discs indicates that the mm-sized particles have rather large Stokes numbers, limiting disc masses to only $\sim 1\mj$. Furthermore, hydrodynamical simulations and population synthesis show that planets with properties deduced from these observations should both grow in mass and migrate inward very rapidly \citep{ClarkeEtal18,MentiplayEtal18,LodatoEtal19,NayakshinEtal19,NduguEtal19}. This would make the detection of these planets statistically very unlikely; this paradox is resolved if the disc masses are $\sim 1\mj$ so that planets do not migrate rapidly \citep{Nayakshin20-Paradox}.

Here we focus on the protoplanetary disc in TW Hydra \citep{Kastner97-TW-Hya}. At the distance of about 60~pc, this protostar is the closest one with a protoplanetary disc, and is probably the best studied. Despite its advanced age of $t_* \approx 10$~Myr \citep{WeinbergerEtal13}, TW Hydra continues to accrete gas at a respectable rate ($\sim 2\times 10^{-9}\msun$~yr$^{-1}$). Its disc is the prototype for discs with cavities possibly carved by growing planets, with early observations indicating gaps on the sub-AU to a few AU scales \citep{CalvetEtal02,Eisner-05-TWHya}. Recent high resolution ALMA observations found several axisymmetric gaps in the continuum sub-mm dust emission of TW Hya, one on the scale of $\sim 1$~AU, and two more at $\sim 24$ and 41 AU \citep{AndrewsEtal16,HuangEtal16,TsukagoshiEtal16,Dsharp2}. 

However, the currently unique feature of TW Hydra is the first ever image of a potential planet in the ALMA 1.3 mm dust continuum emission  \citep[][T19 hereafter]{TsukagoshiEtal19}. The few AU-scale emission excess is significant ($12 \sigma$ over the disc background intensity) and has an azimuthally elongated  shape. T19 associated it with a circumplanetary disc of a growing Neptune mass planet. Most curiously the putative planet is located not inside the previously discovered gaps but at 51.5 AU from the star, right on the edge of a well known cliff-like rollover in the dust disc \citep{AndrewsEtal12-TW-Hya,AndrewsEtal16,Hogerheijde16-TWHYA}. In contrast, emission of many molecular tracers and from microscopic dust have broad peaks at $50-60$~AU, and extending to distances as large as $100-250$~AU \citep{KastnerEtal15,BerginEtal16,Teague18-TWHya}. 

Below we model the time evolution of the gas and dust components of TW Hydra's disc and compare the resulting dust continuum emission with ALMA bands 7 and 6 ($820\mu$m and 1.3 mm, respectively) and EVLA 9 mm observations of the source. We investigate several possible scenarios to try and explain  both the disc morphology, with its sharp rollover of the dust disc, and the image of a putative planet positioned at the edge of that rollover. We find multiple lines of argument that challenge the standard quasi steady-state scenario for this disc. We propose that this disc has been recently re-invigorated by mass injection from a disrupted planet.

The paper is structured as follows.  In \S \ref{sec:Obs} we summarise the observations of TW Hydra relevant to this paper. In \S \ref{sec:analytics} we give simple analytical arguments that challenge the quasi steady-state scenario. In \S \ref{sec:methods} we present our numerical methods. In \S \ref{sec:steady} we apply these to the quasi steady state disc scenario, finding significant and additional to \S \ref{sec:analytics} problems with it. We test a phenomenological Dust Source model in \S \ref{sec:dust-source}, in which a low-mass object on a fixed circular orbit ejects dust into the surrounding, initially dust-free, disc. This scenario proves promising but is not physically self-consistent. 
In \S \ref{sec:Destroyed-CA-planet} we propose that a collision between a $\sim (30-40)\mearth$  pre-collapse Core Accretion planet with a dust-rich  envelope and another massive core could unbind the envelope and make the planet a dust source. In \S \ref{sec:GI-planet-disruption} we show that a gas giant planet formed in the Gravitational Instability scenario could be disrupted from within by its massive luminous core provided that the planet is very metal rich ($Z\sim  0.1$) and its mass is $M_{\rm p}\lesssim 2 \mj$. 

Since the planet is formed in this scenario very early on (presumably at $t\lesssim 0.1$~Myr), survival of the planet on a very wide orbit by $t\sim 10$~Myr requires that TW Hydra had neither gas nor dust disc before the disruption of the planet. In \S \ref{sec:dust-near-T19} we tackle the issue of the "blob" morphology in the \cite{TsukagoshiEtal19} ALMA image, e.g., its spatial extent and its elongation along the azimuthal direction. We conclude with an extended discussion in \S \ref{sec:Discussion}.





\section{Dusty puzzles of TW Hydra}\label{sec:Obs}

\begin{figure*}
\includegraphics[width=0.95\textwidth]{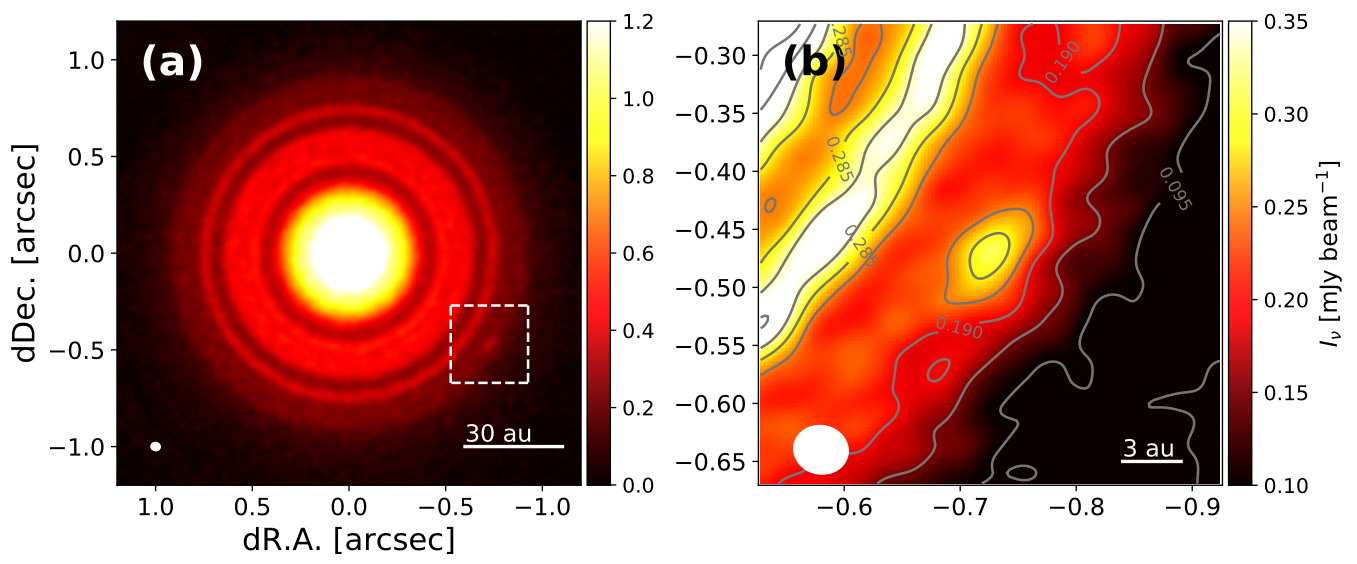}
\includegraphics[width=0.99\textwidth]{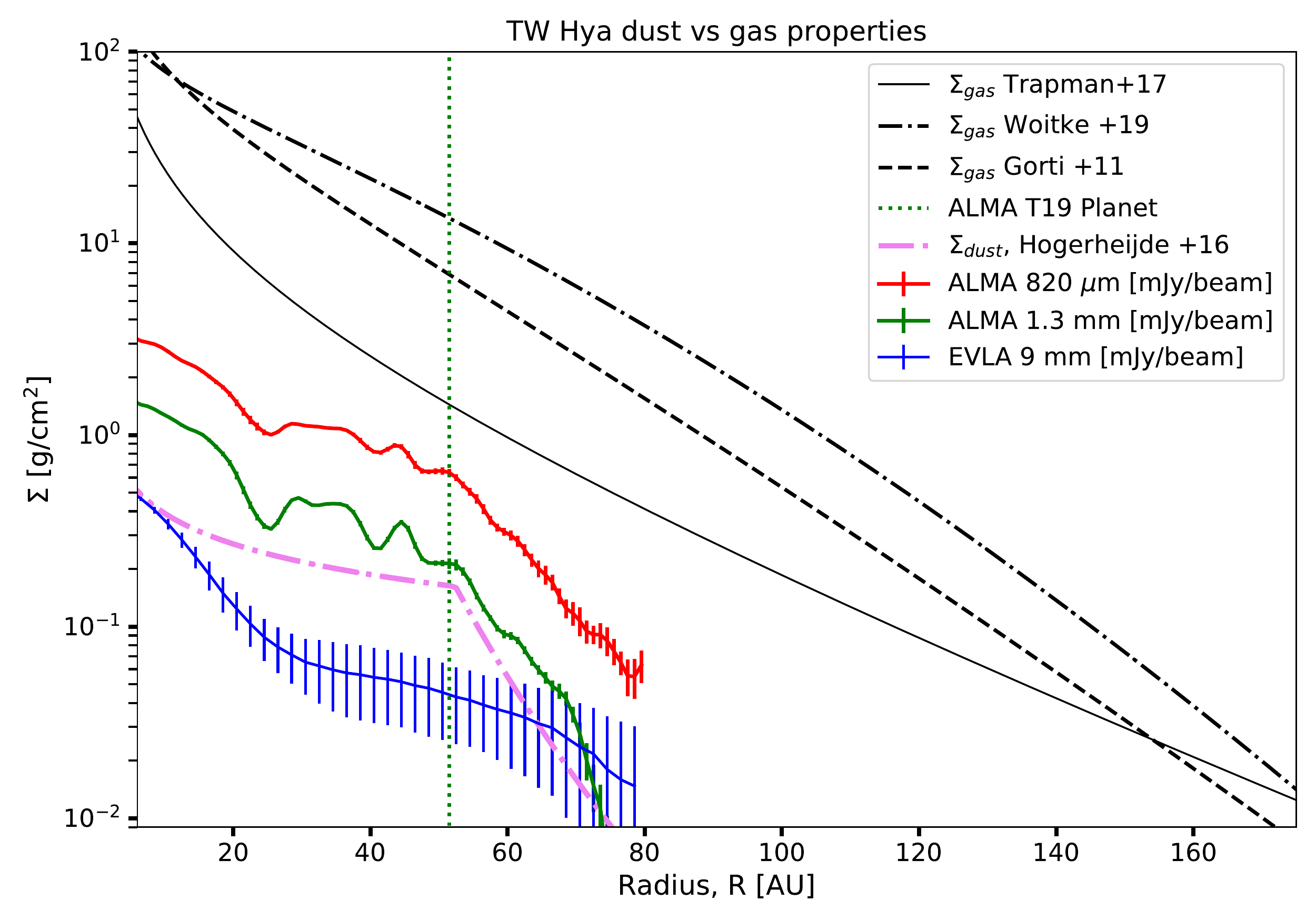}
\caption{{\bf Top:} TW Hydra ALMA image reproduced from Tsukagoshi et al 2019. Note the excess emission inside the white box in panel (a), and the zoom in on this feature in panel (b). {\bf Bottom:} The gas surface density models (black curves) vs azimuthally averaged deprojected ALMA and EVLA dust continuum intensity profiles, and the dust surface density model from Hogerheijde et al 2016. Note how compact the dust distribution is compared to that of the gas, and that the cliff-like rollover in $\Sigma_{\rm dust}$ neatly coincides with the location of the T19 excess emission in the top panels.}
\label{fig:Obs}
\end{figure*}

TW Hydra is an oddity in many regards. The mass of the mm-sized dust in TW Hya is surprisingly large, estimated at $ \sim 80\mearth$ \citep[after correcting to the new GAIA distance;][]{AndrewsEtal16}. \cite{WilliamsEtal19-ODISEA} shows that the mean dust mass of class I sources (generally thought to be much younger than 1 Myr) is $\sim 4\mearth$, whereas the mean mass of class II sources is $\sim 0.8\mearth$. Thus, despite being older by a factor of a few than an average type II disc, TW Hya's dust disc is $\sim 100$ times more massive in dust than the mean\footnote{We shall see later  that TW Hydra's dust mass may be as small as $\sim 15\mearth$ if DIANA \citep{Woitke16-DIANA} opacities (\S \ref{sec:methods-spectrum}) are used. \cite{WilliamsEtal19-ODISEA} $ 1.3$~mm opacity is quite similar to that used by \cite{AndrewsEtal16}. The higher DIANA opacities would reduce the dust mass estimates made by \cite{WilliamsEtal19-ODISEA} by about the same factor, maintaining the surprising mass superiority of TW Hydra's dust disc.}.

The mass of H$_2$ gas disc in TW Hya is suspected to be very high, although there are significant uncertainties. \cite{BerginEtal16} inferred TW Hya H$_2$ disc mass from HD line observations to be $\gtrsim 0.05\msun$ but this value was revised to between $0.006 - 0.009 \msun$ by \cite{TrapmanEtal17} with an alternative model and more data. However, as shown by \cite{JPA12}, for a disc evolving viscously, its mass at age $t_*$ is given by $M_{\rm d} \approx \xi_d \dot M_* t_* \approx 0.1 \msun (\xi_d/5)$, where $\xi_d \gtrsim $ a few. Here the accretion rate inferred for TW Hya is $\dot M_* = 1.8\times 10^{-9} \msun$~yr$^{-1}$ \citep{Ingleby-TTstars-mdots}. Independently of this argument, \cite{Powell19-DustLanes} use the dust radial drift constraints to estimate the gas disc mass in TW Hya as $\sim 0.11\msun$\footnote{In fact the authors assumed the age of $5$~Myr. For a 10 Myr source the required disc mass to delay the dust drift sufficiently would be $\sim 0.2\msun$}. Such a high mass is $\sim 2$ orders of magnitude higher than the median {\em gas} disc mass for type II discs \citep{ManaraEtal18}.

Fig. \ref{fig:Obs} shows the summary of TW Hydra observations that are relevant to this paper. The top left panel shows the image of the source in the 1.3 mm continuum presented in \cite{TsukagoshiEtal19}. The two annular gaps are well known from the previous ALMA observations \citep{AndrewsEtal16,TsukagoshiEtal16,Dsharp2}. The white box in the left panel of Fig. \ref{fig:Obs} is centred on the location of the excess. The half widths of the excess are $\sim 0.5$~AU in the radial and $\sim 2.2$~AU in the azimuthal directions. 

The bottom panel of fig. \ref{fig:Obs} shows with the green and red curves the ALMA azimuthally averaged dust continuum intensity profiles at 1.3 mm and $820\mu$m, respectively, and the EVLA 9 mm intensity with the blue curve\footnote{The data are downloaded from Disks at EVLA program: \url{https://safe.nrao.edu/evla/disks/}.}. The radial profiles of the ALMA images were extracted by averaging the full azimuthal angle except for the position angles of the millimeter blob (P.A.=$222^\circ -236^\circ$ and $238^\circ-252^\circ$). For the profile of the EVLA data, we took an average over the full azimuthal angle. The profiles were made after the image deprojection under the assumption that the disk inclination is $7^\circ$ and the position angle is $-30^\circ$ from the North. We see that the ALMA continuum emission plunges by an order of magnitude from $R\sim 52$~AU to $R=70$~AU. This behaviour is not apparent in the EVLA data, but because the beam size is $\sim 15$~AU it is possible that the intrinsic disc $I_\nu$ also has a similarly sharp rollover at the same location.

While dust discs much smaller than the gas discs as traced by CO emission are not uncommon, and while there are other dust discs with sharp outer cutoffs \citep[e.g.,][]{Trapman19-Dust-vs-Gas}, TW Hya is certainly the best resolved disc of this kind. \cite{Hogerheijde16-TWHYA} searched for a broken power-law fit to 
 ALMA data at the wavelength of $820 \mu$m. They found the power-law index of $p \approx -0.5$ before the break, and  $\Sigma_{\rm dust} \propto R^{-8}$ beyond the break. We show this $\Sigma_{\rm dust}(R)$ profile with the dot-dash violet curve in Fig. \ref{fig:Obs}. Note that the location of the T19 planet coincides very well with the break in dust $I_\nu$ and $\Sigma_{\rm dust}$. The gas surface density profile is much harder to constrain although it is clear that the gas disc extends at least to 200 AU. The three black curves in the bottom panel of fig. \ref{fig:Obs} show three gas surface density models from the literature. While the total gas disc mass varies by about an order of magnitude between these, it is clear that the gas distribution is not as compact as that of the dust.

 \section{Analytical preliminaries: the steady-state scenario}\label{sec:analytics}

In the classical protoplanetary disc paradigm, the discs are formed during the collapse of the parent molecular cloud and persist until dispersed after $\sim (3-10)$~Myrs \citep{AlexanderREtal14a}. This Steady State scenario is explored numerically in \S \ref{sec:steady}, here we scrutinise it with more transparent analytical arguments. 


\subsection{Disc model}\label{sec:disc-model}

The accretion rate onto the star in the viscous steady state is
\citep{Shakura73}
\begin{equation}
    \dot M_* = 3\pi \nu \Sigma = 3\pi \alpha\left(\frac{H}{R}\right)^2 v_K R \Sigma\;,
    \label{Mdot-alpha-0}
\end{equation}
where $v_K = \Omega_K R = (GM_*/R)^{1/2}$ is the local circular Keplerian speed,  $\Sigma$ is the local gas surface density, $\nu = \alpha c_s H$ is the disc viscosity, $\alpha$ is the viscosity parameter, $H$ is the disc vertical scale height, and $c_s$ is the gas sound speed. We assume that the disc midplane temperature is given by
\begin{equation}
    T = T_0 \left(\frac{R_0}{R}\right)^{1/2}\;,
    \label{TR0}
\end{equation}{}
where $T_0 = 18$~K is the gas temperature at the planet location, $R_0 = 51.5$~AU \citep[this temperature profile is very similar to the midplane temperature profile derived in][]{HuangEtal16}. For reference, $v_K = 3.7$ km/s, $c_s = 0.25$ km/s, and $H/R \approx 0.07$ at $R=R_0$. In the whole of \S \ref{sec:analytics} we also assume that 
\begin{equation}
    \Sigma(R) = \Sigma_0 \frac{R_0}{R}\;.
    \label{Sigma-gas0}
\end{equation}{}
Equation \ref{Mdot-alpha-0} constrains the product of $\alpha \Sigma_0$. We shall use additional considerations to constrain $\Sigma$, which then limits $\alpha$ via eq. \ref{Mdot-alpha-0}.

\subsection{Gas accretion constraints on the disc mass}\label{sec:gas_accretion}

Numerical integrations of viscously evolving discs show that the accretion rate onto the star is large at early times and then decreases \citep{JPA12}. The product of the accretion rate and current time $t$ remains approximately constant initially and later also decreases (because $\dot M_*$ drops with time). Usefully for us here, the disc mass at age $t_*$ is given by 
\begin{equation}
    M_{\rm d} \approx \xi_d \dot M_* t_* \approx 0.1 \;\msun\; \left(\frac{\xi_d}{5}\right)\;,
    \label{Mdisc-visc0}
\end{equation}{}
 where $\xi_d \sim $ a few is a dimensionless number \citep{JPA12}, and we used
 $\dot M_* = 1.8\times 10^{-9} \msun$~yr$^{-1}$ and $t_* = 10$~Myr.

\subsection{Dust drift constraints}\label{sec:d-drift}

As previous authors \citep{AndrewsEtal12-TW-Hya} we find that particles of size $a \gtrsim 1$~mm are required to explain ALMA observations. The radial dust drift velocity is
\citep{Whipple72}
\begin{equation}
    v_{\rm dr} = \eta \left(\frac{H}{R}\right)^2 \frac{ v_{\rm K} }{\St + \St^{-1}}\;,
    \label{v-drift0}
\end{equation}{}
where $\eta = 1.25$ given our model disc pressure profile, and the Stokes number is given by 
\begin{equation}
    \St = \frac{\pi}{2} \frac{\rho_a a}{\Sigma}\;,
    \label{Stoke0}
\end{equation}{}
with $\rho_a =2.1$~g cm$^{-3}$ \citep{WoitkeEtal19} being the grain material density. For a $p=3.5$ power-law grain size distribution with maximum grain size $a$, the appropriate grain size to use in eqs. \ref{v-drift0} \& \ref{Stoke0} in the small $\St \ll 1$ regime is the mean grain size, $a_{\rm mean} = a/3$ (see \S \ref{sec:grain-growth}). Demanding the dust particle drift time scale to be 10 Myr at the location of T19 planet we arrive at the minimum gas surface density at $\sim 50$~AU of
\begin{equation}
    \Sigma_{\rm 0} > 100 \hbox{ g cm}^{-2} \; \frac{\hbox{1 mm}}{a}\;.
    \label{Sigma-min0}
\end{equation}{}
With the observed gas accretion rate of $\dot M_*\approx 1.8 \times 10^{-9} \msun$~yr$^{-1}$, we simultaneously have that
\begin{equation}
    \alpha <  10^{-4}\;.
    \label{alpha-min0}
\end{equation}{}
The disc mass enclosed within radius $R$ is then
\begin{equation}
    M_{\rm disc}(R) = 2\pi R R_0 \Sigma_0 = 0.19 \msun \frac{R}{R_0}
    \label{Md-min-drift}
\end{equation}{}
This is larger than $M_{\rm d} = 0.11\msun$ found by \cite{Powell19-DustLanes}, mainly because they assumed that TW Hydra is younger (5 Myr). 

The mass in eq. \ref{Md-min-drift} is uncomfortably large for many reasons.  Firstly, the typical gas mass of a $\sim$ few Myr old disc is estimated at $\sim 10^{-3}\msun$ \citep{ManaraEtal18}. Secondly, such a massive disc should be strongly self-gravitating since the \cite{Toomre64} parameter is
\begin{equation}
    Q = \frac{c_s \Omega}{\pi G\Sigma} \sim 0.6
    \label{Q0}
\end{equation}{}
when evaluated at $R = R_0$. We expect the disc to show spiral density structure, and in fact be fragmenting for such a low $Q$. Indeed, a 3D Phantom SPH calculation confirmed that the disc with the structure introduced at \S \ref{sec:disc-model} and extending to 200 AU (as observed) fragments due to self-gravity.

Another interesting inference from the dust drift constraints is the dust-to-gas ratio in the TW Hydra disc. According to \cite{Hogerheijde16-TWHYA}, $\Sigma_{\rm dust} \approx 0.2$ at $R\sim 50$~AU (although this depends on the dust opacity and size distribution). Hence, $\varepsilon_0 = \Sigma_{\rm dust}/\Sigma \approx 0.003$.
This is only slightly larger than that derived by \cite{WoitkeEtal19} for TW Hydra, who obtained $\varepsilon_0 \approx 0.002$. If the observed emission of the T19 feature is due to a circumplanetary disc around the planet, we would then expect the disc to have a dust-to-gas ratio smaller than $\varepsilon_0$ because dust in the circumplanetary disc is expected to drift into the planet faster than the gas component of the disc does (similarly to the protoplantary disc case). Therefore, we conclude that the minimum gas mass of the circumplanetary disc (CPD) is
\begin{equation}
    M_{\rm cpd} > M_{\rm cpd-dust} \varepsilon_0^{-1} = 10 \mearth\;,
    \label{M-cpd0}
\end{equation}{}
where $M_{\rm cpd-dust} \approx 0.03\mearth$ is the dust mass of the T19 feature \citep{TsukagoshiEtal19}. This CPD mass is surprisingly high and would require the planet itself to be much more massive, e.g., many tens of $\mearth$ to maintain dynamical stability. This high $M_{\rm p}$ in combination with high $M_{\rm d}$ is ruled out due to planet migration (\S \ref{sec:typeI-migr}) and spectral constraints (\S \ref{sec:steady}).

\subsection{Dust particles maximum size}\label{sec:d-size}


Here we discuss three processes that limit grain growth in our numerical models below. We also use these results to argue that the quasi steady state scenario cannot naturally explain the relatively small grain size in this old disc (cf. further \S \ref{sec:SS-discussion}, item \ref{item:very-low-vbreak}).


\cite{BirnstielEtal12} conclude that dust particle collisions due to radial drift are not likely to be a dominant mechanism of dust size regulation. For the steady state scenario of TW Hydra disc in particular, the drift velocity can be estimated by requiring the dust to not drift all the way into the star in 10 Myr:
\begin{equation}
    v_{\rm dr} \lesssim \frac{R_0}{\hbox{10 Myr}} \approx 0.025 \hbox{ m s}^{-1}\;.
    \label{vdr-ss0}
\end{equation}
Grains of similar sizes collide at $\sim 1/2$ of this velocity, e.g., at just about 1 cm~s$^{-1}$. This velocity is two orders of magnitude smaller than the grain breaking velocities inferred from experiments and typically considered in the field, $v_{\rm br}\sim 10$~m s$^{-1}$. Note however that this process can be dominant for planet-disruption scenarios  where the relevant time scale is $\sim 10^5$~yr.


The grain growth time is finite. \cite{BirnstielEtal12} introduce the "drift limit" to the maximum grain size as the grain size to which the grains grow before they are efficiently removed by the radial drift: 
\begin{equation}
    a_{\rm drift} = 0.55 \frac{2}{\pi} \frac{\Sigma_{\rm dust}}{\eta \rho_a} 
    \left(\frac{R}{H}\right)^2 = 54 \hbox{ mm}
    \label{adrift0}
\end{equation}{}
at the location of the T19 planet (we used here $\Sigma_{\rm dust} = 0.2$ g/cm$^2$ from \cite{Hogerheijde16-TWHYA}). This value is quite large, indicating that this process is also unlikely to stem grain growth in TW Hydra.


Finally, gas turbulence also sets a maximum grain size \citep{Weidenshilling84}, which in the limit of small Stokes number yields 
\begin{equation}
    a_{\rm turb} = f_{\rm t} \frac{2}{3\pi} \frac{\Sigma}{ \rho_a\alpha_{\rm t}} \frac{v_{\rm br}^2}{c_s^2}\;.
    \label{a-turb0}
\end{equation}
Here we used the turbulent viscosity parameter $\alpha_{\rm t}$, which may in general be different from the \cite{Shakura73} $\alpha$ viscosity parameter introduced previously. The latter parameterizes the efficiency of the angular momentum transfer and subsumes in itself both gas turbulence and the effects of possible large scale magnetic torques \citep{BaiStone13,Bai16}. Since $\alpha \ge \alpha_{\rm t}$, we estimate the minimum size set by turbulence via setting $\alpha_{\rm t} = \alpha$:
\begin{equation}
    a_{\rm turb} \gtrsim 6 \hbox{ mm} \left(\frac{v_{\rm br}}{1 \hbox{ m s}^{-1}}\right)^2 \alpha_{-4}^{-1}\frac{\Sigma_0}{ 100 \hbox{ g cm}^{-2}}
    \label{a-turb1}
\end{equation}{}
where $\alpha_{-4} = 10^4 \alpha$ (cf. equation \ref{alpha-min0}). For the typical values of $v_{\rm br} \sim 10$ m~s$^{-1}$ employed in the literature \citep[e.g.,][]{BirnstielEtal12}, grains should grow much larger than $\sim 1$~mm and then drift inward too rapidly (cf. \S \ref{sec:d-drift}). Since the observed gas accretion rate on TW Hydra sets the constraint $\alpha \Sigma_0 = $~const (cf. \S \ref{sec:disc-model}), we see that the scaling in eq. \ref{a-turb0} is 
\begin{equation}
    a_{\rm turb} \propto v_{\rm br}^2 \Sigma_0^2\;.
    \label{sec:a-turb1}
\end{equation}{}
This shows that a less massive disc would naturally result in more reasonable maximum grain sizes, avoiding the unwelcome necessity to demand that $v_{\rm br}$ is as small as $\sim 0.2$~m/s.

\subsection{The sharp dust rollover in TW Hydra}\label{sec:SS-the-drop}

TW Hydra displays a cliff-like rollover in the dust density distribution at separation $R \simeq 52$~AU \citep{AndrewsEtal12-TW-Hya,Hogerheijde16-TWHYA,TsukagoshiEtal19}. The observed rollover can be fit with a power-law $\Sigma_{\rm dust} \propto R^{-8}$ at $R \gtrsim 50$~AU \citep{Hogerheijde16-TWHYA}. In the context of the standard paradigm for protoplanetary discs, the separation of the dust and the gas may be expected due to the radial drift of the dust and the viscous spreading of the gas disc \citep[e.g.,][]{Powell19-DustLanes,Rosotti19-Opacity-Cliff,Trapman19-Dust-vs-Gas}.

Let us consider the dust radial drift time scale dependence on distance $R$:
\begin{equation}
    t_{\rm dr} = \frac{R}{v_{\rm dr}} \propto \frac{R\Sigma}{a(R)}\;,
\end{equation}{}
where $a(R)$ is the radius-dependent grain size, and we assumed the low Stokes number limit. Steady-state discs usually have $\Sigma R \approx $~const, as we also assumed in eq. \ref{Sigma-gas0}. Thus a decreasing grain size with increasing $R$ means that $t_{\rm dr}$ increases with $R$.

This in turn implies that dust density gradients are erased over time. Consider two radii in the disc, $R_1 < R_2$, and $\Sigma_1 > \Sigma_2$. Since the drift time scale at $R_1$ is shorter than that at $R_2$, the dust surface density at $R_1$ drops with time faster than it does at $R_2$, and hence the ratio $\Sigma_1/\Sigma_2$ decreases with time. 

Therefore, if the TW Hydra dust distribution does evolve from some initial distribution, then that distribution must have had an even steeper rollover than the currently observed $\propto R^{-8}$ one. It seems rather contrived to demand  such a sharp initial dust edge at time $t\approx 0$.

\subsection{Planet migration constraints}\label{sec:typeI-migr}

The type I migration time scale at the location of T19 planet is very short:
\begin{equation}
    t_{\rm mig1} = \frac{1}{2\gamma \Omega} \frac{M_*^2}{M_{\rm p} \Sigma R^2} \left(\frac{H}{R}\right)^2 \approx 4\times 10^4 \hbox{ yr } \frac{ 100 \hbox{ g cm}^{-2}}{\Sigma_0} \frac{10 \mearth}{M_{\rm p}}
    \label{t1-T19}
\end{equation}
where the dimensionless factor $\gamma$ \citep{PaardekooperEtal10a} evaluates to $\gamma=2.5$ in our disc model. The planet migration time is less than 1\% of TW Hydra's age. Assuming that we are not observing the system at a special time, TW Hydra should hatch $\sim 100$ such planets over the course of its protoplanetary disc lifetime for us to have a decent statistical chance to observe it. This does not appear reasonable; most likely the gas disc is much less massive.


 \section{Methodology}\label{sec:methods}

We model the time dependent evolution of dust and gas components in a 1D azimuthally averaged viscous disc with an embedded planet that can optionally lose a part of its mass to the disc.

\subsection{Gas and planetary dynamics}\label{sec:dynamics}

Our code builds on the work of \cite{NayakshinLodato12}, who modelled the time-dependent evolution of a viscous gaseous disc in azimuthal symmetry, with the disc interacting with an embedded planet via gravitational torque and optionally exchanging mass. Without the mass exchange, the planet modifies the disc structure near its orbit only via these torques; the reverse torques from the disc onto the planet force it to migrate, usually inward. The corresponding equations for the gas surface density $\Sigma$ are eqs. 36-37 (without the mass exchange term for now), and eq. 45 for the orbital radius evolution (migration) of the planet in \cite{NayakshinLodato12}. We do not solve for the thermal balance of the disc here, assuming that the disc is passively heated and the midplane temperature is given by eq. \ref{TR0}. Our neglect of disc viscous heating is physically reasonable since we consider much larger orbital separations and much smaller stellar accretion rates than did \cite{NayakshinLodato12}. 

If the planet is massive enough, a deep gap in the gas surface density profile opens up due to the planetary torques on the disc, and the planet then migrates in the type II regime \citep[see][]{Nayakshin15c}. We use a logarithmic bin spacing in radius $R$ from $R_{\rm in} = 2$~AU to $R_{\rm out} = 300$~AU with, typically, 250 radial zones. The mass exchange term is not present in the massive quasi-steady disc scenario (\S \ref{sec:steady}) and thus will be discussed later when needed. 

The initial gas surface density profile is given by
\begin{equation}
    \Sigma(R) = \Sigma_0 \frac{R_0}{R} \exp\left( -\frac{R}{R_{\rm exp}} \right)\;,
    \label{sigma-disc0}
\end{equation}{}
where the exponential rollover $R_{\rm exp} = 100$~AU, which is reasonable given the observed radial extent of the CO gas disc.

\subsection{Dust disc evolution}\label{sec:methods-grain-dynamics}

Following \cite{DipierroLaibe17} we extend the code to include the dust component in the disc, although setting $\epsilon = \Sigma_{\rm dust}/\Sigma = 0$ in their relevant equations as we assume the dust to be in the test particle regime. The time-dependent radial drift and turbulent diffusion equation for the dust is
\begin{equation}
    \frac{\partial \Sigma_{\rm d}}{\partial t} + \frac{1}{R}\frac{\partial }{\partial R}\left[R \Sigma_{\rm d} v_{\rm dr}'\right] = \frac{1}{R}\frac{\partial }{\partial R}\left[R D \Sigma \frac{\partial }{\partial R}\left(\frac{\Sigma_{\rm d}}{\Sigma}\right) \right] 
    \label{dSdt0}
\end{equation}{}
where $v_{\rm dr}'$ is the full dust drift velocity given by sum of the radial drift velocity (eq. \ref{v-drift0}), the additional component due to the gas radial flow and the gravitational torque term from the planet \citep[see eq. 16 in][for the full expression]{DipierroLaibe17}. $D$ is the turbulent diffusion coefficient for dust, which is related to the turbulent gas viscosity $\nu = \alpha_{\rm turb} c_s H$ via
\begin{equation}
    D = \frac{\nu}{1 + \St^2}\;,
\end{equation}{}
where the Stoke number is given by eq. \ref{Stoke0}.  

\subsection{Grain size evolution}\label{sec:grain-growth}

Eq. \ref{dSdt0} is designed to follow the evolution of dust particles of a fixed size (Stokes number). It is desirable to extended the method to a distribution of grain particles. It is currently prohibitively expensive to model numerically the grain particle size evolution together with the spatial evolution of grains. A physically reasonable approximation, commonly employed in the literature, is to assume that the dust follows a power-law size distribution at all locations in the disc, $dn/da_g \propto a_g^{-p}$, with $p = 3.5$ for grain sizes between a minimum and a maximum, $a_{\rm min} \le a_g\le a$. The maximum grain size is allowed to evolve in space and time due to grain growth and collisions \citep[cf.][]{BirnstielEtal12,Vorobyov-Elbakyan-19,Rosotti19-Opacity-Cliff}. The minimum grain size is of a little consequence at mm wavelengths \citep[e.g., see the top left panel in fig. 3 in][]{WoitkeEtal19} and so we fix $a_{\rm min} = 0.05\mu$m.  Note that for $p = 3.5$, most of the mass is at the largest sizes of dust particles, and the mean grain size is $a_{\rm mean} \approx a/3$. By following the dynamics of particles with size $a_{\rm mean}$ we then follow the dynamics of the bulk of the dust \citep[as also done, for example, by][]{Rosotti19-Opacity-Cliff}.

We start the simulations with the maximum grain size being small everywhere in the disc, $a(R) = 1\mu$m. We then allow the grains to grow until the maximum grain size reaches either one of the three well known maximum grain size constraints -- the turbulent fragmentation, the radial drain, or the radial drift fragmentation limits \citep{BirnstielEtal12} -- as described in \S \ref{sec:d-size}.

\subsection{Computing the disc emission spectrum}\label{sec:methods-spectrum}

Once we have the dust surface density and the maximum grain size distributions for all disc radii, $\Sigma_{\rm d}$ and $a(R)$, we compute the dust optical depth  
\begin{equation}
    \tau_{\rm d} = \kappa(\lambda, a) \Sigma_{\rm d}(R)\;,
    \label{tau-d0}
\end{equation}{}
where $\kappa(\lambda, a)$ is the DIANA dust opacity \citep{Woitke16-DIANA} computed for the maximum grain size $a$ and radiation wavelength $\lambda$. Both scattering and absorption opacities are included in $\kappa(\lambda, a)$. We used an amorphous carbon fraction of 26\%, higher than the standard value used by \cite{Woitke16-DIANA}. We found this to be necessary to fit the relative luminosities of TW Hydra in 820 $\mu$m and 1.3 mm wavelengths. We subsequently found that this is very close to the 25\% amourphous Carbon fraction derived by \cite{WoitkeEtal19} for TW Hydra.

We compute the disc surface brightness at radius $R$ as
\begin{equation}
    I_\nu(R) = \zeta_s B_\nu\left[T(R)\right] \; \left(1 - e^{-\tau_{\rm d}}\right)\;,
    \label{Emissivity0}
\end{equation}{}
where $T(R)$ is the local disc midplane temperature, and $0 < \zeta_s \le 1$ is the dimensionless function describing the disc emissivity reduction due to dust scattering given by eq. 11 in \cite{Zhu19-scattering-albedo}. As shown by these authors, when $\tau_{\rm d} \ll 1$, eq. \ref{Emissivity0} reduces to the standard optically thin expression \citep[used by, e.g.,][]{AndrewsEtal16} for the radiation intensity emitted by the disc, which has no scattering contribution. However, when $\tau_{\rm d} > 1$, the scattering albedo may produce a non-trivial and significant reduction of the dust emissivity from the blackbody function $B_\nu$; this reduction is described by the function $\zeta_s$. As TW Hydra's disc inclination to us is very small ($i\approx 7^\circ$, $\cos i \approx 0.99$), we shall for simplicity set $i=0$ in this paper.

\begin{figure}
\includegraphics[width=0.48\textwidth]{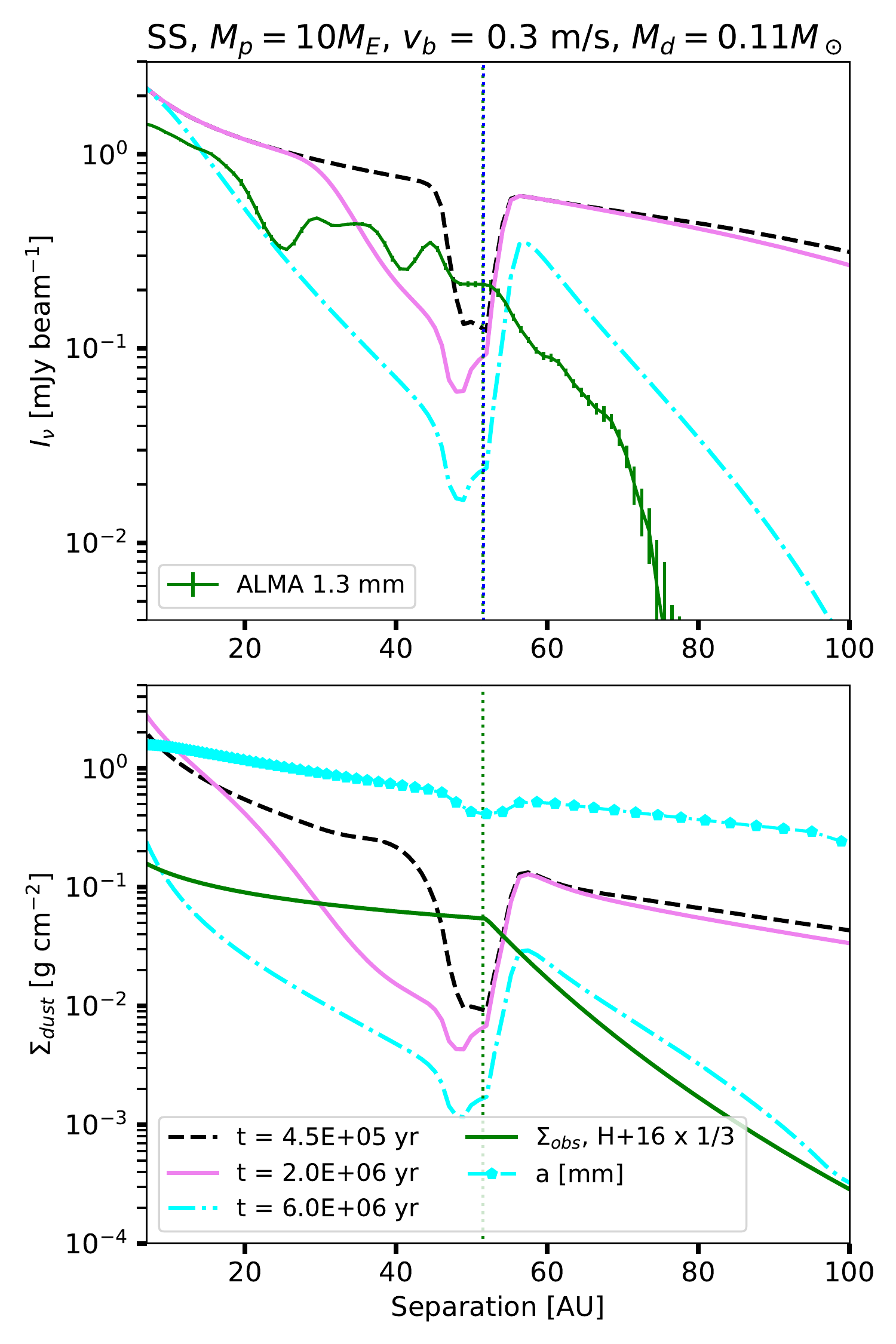}
\caption{The Steady State disc model with an $M_{\rm p} =10 \mearth$ planet located at $51.5$~AU. {\bf Top}: Model disc intensity at three different times as shown in the legend, and the ALMA azimuthally averaged intensity at 1.3 mm (green). {\bf Bottom:} The corresponding dust surface density profiles $\Sigma_{\rm d}$, compared with that inferred from the observations by Hogerheijde et al (2016) (scaled down by a factor of 3 due to different opacities).}
\label{fig:SS-M10}
\end{figure}

\section{The Quasi Steady State disc scenario}\label{sec:steady}

As shown in \S\S \ref{sec:gas_accretion} and \ref{sec:d-drift}, the gas disc must be be very massive to both feed TW Hya and prevent the mm-sized dust drifting into the star in 10 Myr. We set the initial disc mass to $0.15 \msun$ to avoid it becoming self-gravitating. For this disc we found that the viscosity parameter of $\alpha = 10^{-4}$ yields stellar accretion rate between $(1-2)\times 10^{-9}\msun$~yr$^{-1}$ at time between $\sim 3$ and 10 Myr, as appropriate for TW Hya.

For simplicity, we  {\em artificially} hold a planet on a fixed orbit at the location of excess emission in T19. For planets with mass $M_{\rm p}\sim 10\mearth$, the planet migration time scales are uncomfortably short for massive gas discs, $t_{\rm mig}\lesssim 0.1$~Myr (see eq. \ref{t1-T19}). We explored the parameter space of such more self-consistent models and found that they are challenged by the data even more than the fixed planet orbit models. For brevity we do not show their results here.

\subsection{10 Earth mass planet at 51.5 AU}\label{sec:10ME-planet}

Here we present a simulation with the following parameters: an initial disc mass $M_{\rm d} = 0.11\msun$ \citep[as suggested for TW Hydra by][]{Powell19-DustLanes}, the initial dust to gas ratio of $\Sigma_{\rm dust}/\Sigma = 0.03$, and the planet mass $M_{\rm p} =10\mearth$.  The disc viscosity parameter $\alpha=1.5\times 10^{-4}$ was constrained by demanding the gas accretion rate onto the star to be close to the observed value. The turbulent viscosity parameter $\alpha_{\rm turb}$ was set equal to $\alpha$. Note that lower values of $\alpha_{\rm turb}$ are unlikely based on the analysis in \cite{DSHARP-6}. The maximum grain size in TW Hydra disc is at least 1 mm \citep{AndrewsEtal16}. To achieve this, the dust breaking velocity had to be set much lower in this simulation, $v_{\rm br} = 0.3$~m/s, than the value usually assumed in literature \citep[$v_{\rm br}\sim 10$~m~s${-1}$, e.g.,][]{DrazkowskaEtal14,Rosotti19-Opacity-Cliff}.

Fig. \ref{fig:SS-M10} shows the model disc intensity at wavelength 1.3 mm (top panel) and the dust disc surface density (bottom panel) at three different times. The observed ALMA intensity and the dust surface density profile deduced by \cite{Hogerheijde16-TWHYA} for TW Hydra are shown with the solid green curves in the top and bottom panels, respectively. This deduced $\Sigma_{\rm dust}$ is scaled down by a factor of 3 as the DIANA opacities we use are higher by a factor of a few. The bottom panel also shows the maximum grain size $a$ at time $t=6$~Myrs.

We observe a number of features expected from the presence of a massive planet in a disc \citep{RiceEtal06,PinillaEtal12,DipierroLaibe17,Dsharp7}. The planet acts as a dam for the dust, so that a bright outer rim appears behind it. Inside the orbit of the planet, the dust is free to drift into the star, and hence a deep and wide gap appears there. Note that the disc intensity in the top panel is not simply linearly proportional to the dust surface density from the bottom panel. This occurs because the dust grain sizes vary with location in the disc, and even more importantly because the disc intensity saturates at the Blackbody function $B_\nu(T)$ at very high optical depths $\tau_{\rm d}$. As a specific example, consider radius $R\approx 30$~AU in fig. \ref{fig:SS-M10}. While $\Sigma_{\rm dust}$ at this radius decreased by almost an order of magnitude going from $t=0.45$~Myr to $t=2$~Myr, the intensity of the disc emission at that radius did not vary at all. 

Overall we see that the model disc intensity is very different from the one observed, and the dust surface densities are also different from the broken power-law fit of \cite{Hogerheijde16-TWHYA}. Although the disc intensity and $\Sigma_{\rm dust}$ vary with model parameters, in all of the cases we experimented with the model always contradicts the observations: for a sufficiently massive planet, the dust emission should be suppressed inside the planetary orbit and that there should be a bright rim behind it. The observations show no suppression of the dust emission at or inside the orbit of the planet, and the disc intensity does not display a bright rim behind it.

\subsection{3 Earth mass planet at 51.5 AU}\label{sec:3ME-planet}

\begin{figure}
\includegraphics[width=0.48\textwidth]{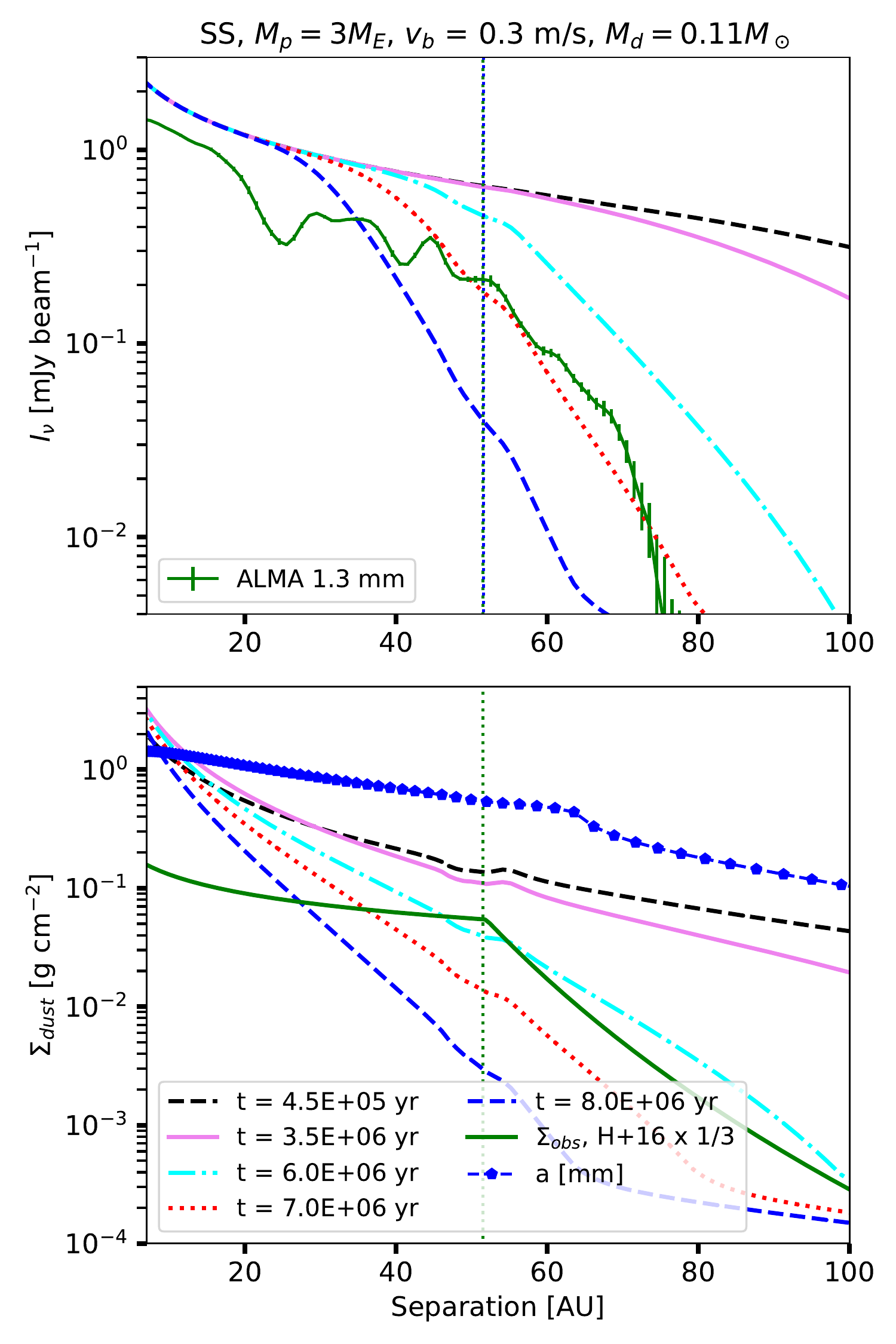}
\caption{Same as fig. \ref{fig:SS-M10} but for planet mass $M_{\rm pl} = 3\mearth$. The effects of the planet on the disc are now barely observable just around its orbit.}
\label{fig:SS-M3}
\end{figure}

Fig. \ref{fig:SS-M3} shows a calculation entirely analogous to that shown in fig. \ref{fig:SS-M10}, except the planet mass is set at $M_{\rm p} = 3\mearth$. In this case the planet produces only a barely noticeable depression in dust surface density just around its orbit, as the observations demand. The results of this calculation are very similar to that with any smaller planet mass, $0 \le M_{\rm p} \le 3 \mearth$.

\begin{figure}
\includegraphics[width=0.48\textwidth]{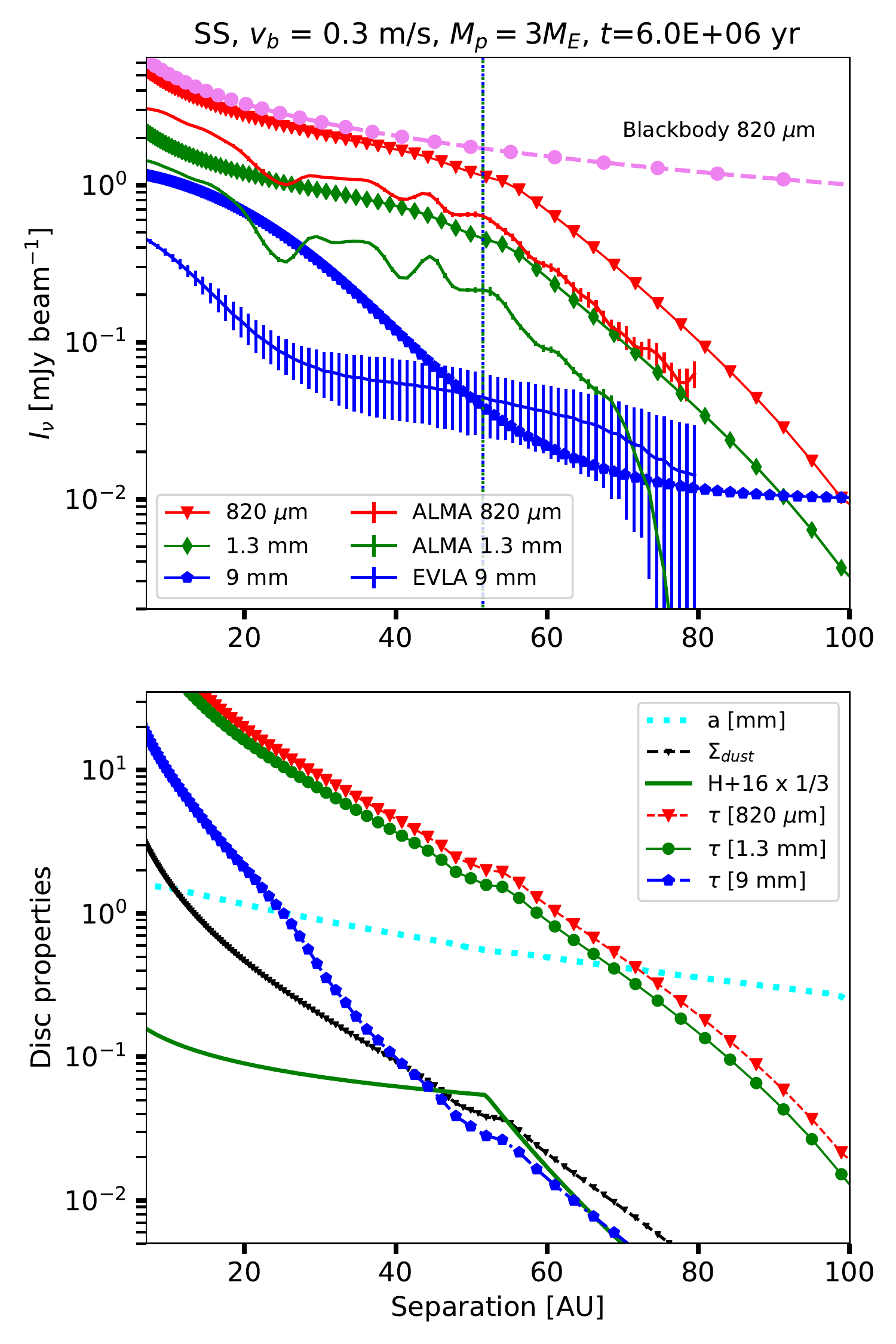}
\caption{Disc Intensity profiles for the three wavelengths (top panel) and the disc properties (lower panel) for the $t=6$~Myr dust profile from fig. \ref{fig:SS-M3}. Note that while this model matches the location and shape of the rollover well, it over-predicts the disc luminosity by a factor of $\sim 3$ in the ALMA bands, as well as contradicting the EVLA data strongly. }
\label{fig:SS-M3-3Wavelengths}
\end{figure}

Fig. \ref{fig:SS-M3} shows that, with the planet effects on the disc reduced, the model may actually yield a flat emissivity profile in the inner disc followed by a steep decline, exactly as needed to explain TW Hydra's ALMA data. This occurs due to the already mentioned saturation of the intensity in the inner disc where it becomes optically thick. For example, the dash-dotted cyan curves ($t=6$ Myr) in fig. \ref{fig:SS-M3} appears most promising, with the break in the disc intensity occurring right where needed. However, the saturation of disc intensity at the Blackbody function is also the reason why this scenario contradicts the data strongly. 

Fig. \ref{fig:SS-M3-3Wavelengths} shows the disc intensity at $t = 6$~Myr in three wavelengths in the top panel, along with the corresponding ALMA and EVLA observations. The bottom panel shows the respective disc scattering plus absorption optical depth for these wavelengths, the dust surface density $\Sigma_{\rm dust}$ (which is the same as the cyan line from the bottom panel in fig. \ref{fig:SS-M3}), and the maximum grain size $a$. While the model fits the broken power-law {\em shape} of the intensity profile of the disc in the two ALMA bands, it is too bright by a factor of $\sim 3$. This cannot be "fixed" by any changes in the dust opacity model or variations in $\sigma_{\rm dust}$. To understand why, note the purple curve in the top panel of fig. \ref{fig:SS-M3-3Wavelengths} that shows the disc intensity profile in the optically thick limit, i.e., $I_\nu = B_\nu$ everywhere. We now see that the break in the disc emissivity profile in the two ALMA bands indeed occurs where its optical depth exceeds unity somewhat (where the dust {\em absorption only} optical depth is $\sim 1$). 

The only physical way to make the model disc appropriately bright in the ALMA bands is to demand that it becomes optically thick not at $R\sim 50$~AU but at $R\sim 30$~AU. However, that would contradict the observed intensity profile. Further, 9 mm EVLA data pose a separate but physically similar challenge. To match the correct intensity level at $\sim 50$~AU in this wavelength the disc needs to be very optically thin. This then implies that the EVLA emission must track the strong rise in $\Sigma_{\rm dust}$ inward, but the observed profile is rather flat in the $\sim 30-60$~AU region.

In fact, it is well known that the outer disc in TW Hydra must be optically thin in sub-mm and longer wavelengths from pre-ALMA photometry (the integrated disc luminosity) data: the luminosity of the source rises as $\propto \nu^{2.6}$ rather than $\propto \nu^2$ as expected for the optically thick disc \citep[e.g., see \S 3 and fig. 1 in][]{Pascucci12-Photo-evap}. Furthermore, the image of the T19 excess emission is significantly smaller than the disc pressure scale height, and that too implies that the disc is optically thin in the ALMA bands (see \S \ref{sec:disk-is-opt-thin}).

Summarising, the quasi steady state scenario cannot fully account for the observed continuum dust emission from TW Hydra's protoplanetary disc. In \S \ref{sec:dust-near-T19} \& \S \ref{sec:SS-discussion} we shall see that it faces many additional challenges.

\section{A phenomenological Dust Source model}\label{sec:dust-source}

We now make a single but significant alternation to the physical setup of our simulations. We assume that the planet ejects dust in the surrounding disc. The simulation setup and initial conditions are exactly the same as those presented in \S \ref{sec:3ME-planet} except that we assume a negligible amount of dust into the gaseous disc at $t=0$ for simplicity. The dust mass loss rate from the planet is a free parameter of the model; we found that choosing $\dot M_Z = 6 \times 10^{-6} \mearth$~yr$^{-1}$ (with other parameters of the model unchanged) provides a somewhat promising spectral model. As in \S \ref{sec:3ME-planet} we keep the planet mass and orbital radius fixed for now, even though this violates both mass conservation and Newton's second law.

\begin{figure}
\includegraphics[width=0.48\textwidth]{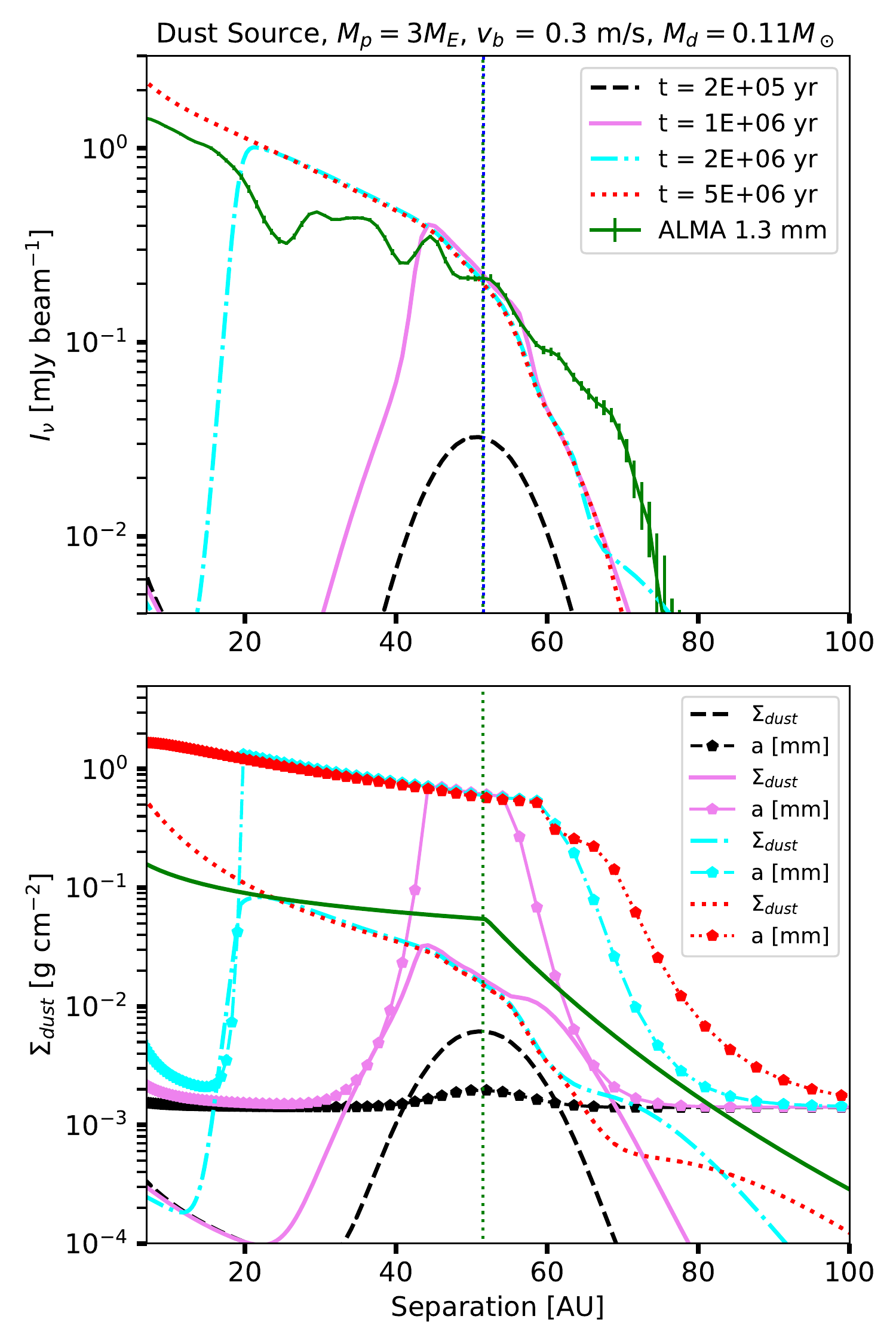}
\caption{Same as fig. \ref{fig:SS-M3} but now for a Dust Source model. The planet $M_{\rm p} = 3 \mearth$ is fixed at 51.5~AU and ejects dust into the surrounding disc at rate $\dot M_Z = 6\times 10^{-6}$~yr$^{-1}$.}
\label{fig:SS-Inject}
\end{figure}

\begin{figure}
\includegraphics[width=0.48\textwidth]{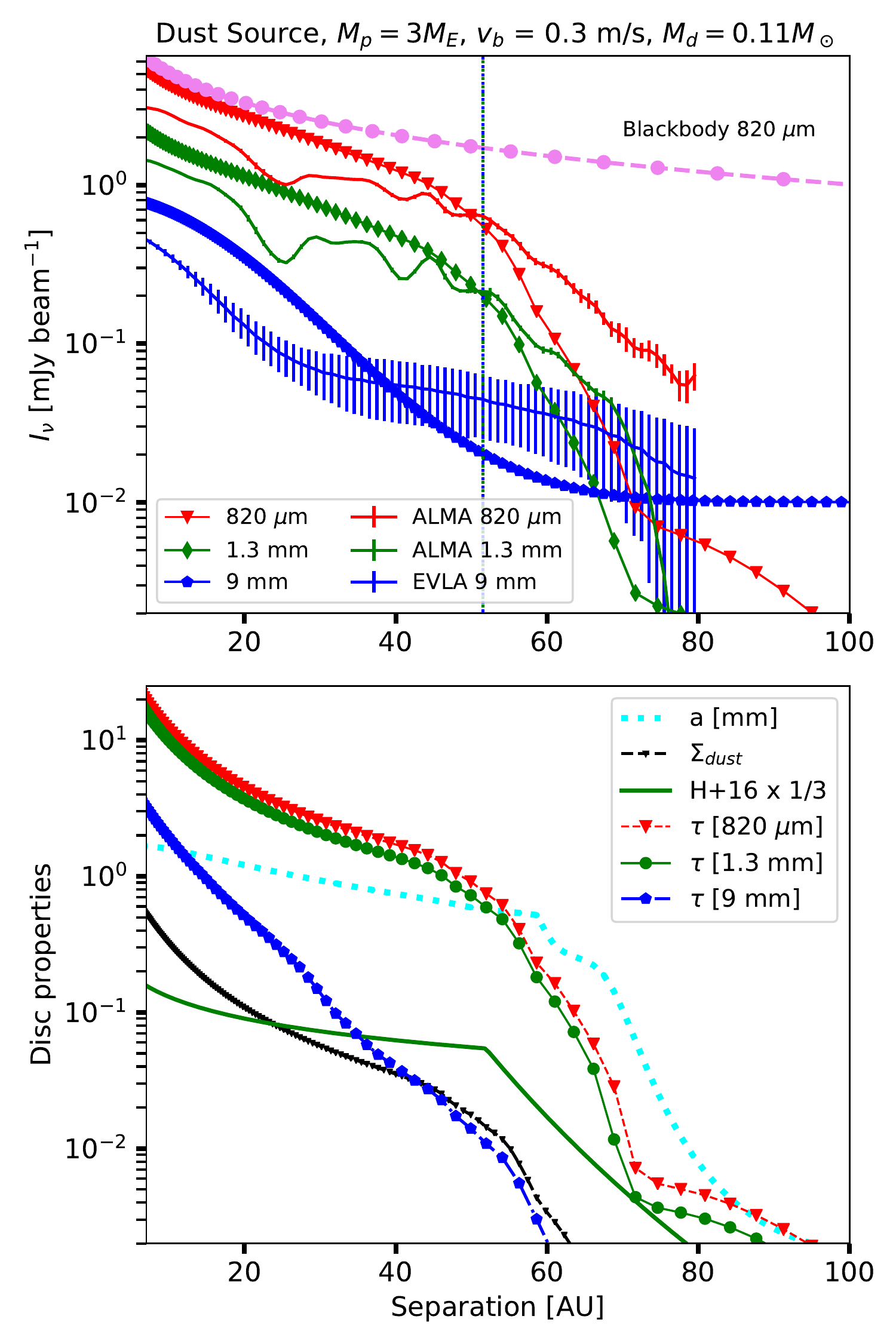}
\caption{Same as fig. \ref{fig:SS-M3-3Wavelengths} but now for the Dust Source model at $t=5$~Myr.}
\label{fig:SS-Inject-3Wavelengths}
\end{figure}

The dust lost by the planet is deposited in a relatively narrow ring with a Gaussian profile around the planet location, with the surface density deposition rate given by
\begin{equation}
    \dot\Sigma = C \exp\left[
    -\frac{(R-R_{\rm p})^2}{2 w_{\rm inj}^2}\right]\;,
    \label{S-injection0}
\end{equation}{}
where $R_{\rm p}$ is the current position of the planet,  and $w_{\rm inj} = 0.15 R_{\rm p}$ is the width of the Gaussian. The normalisation constant $C$ ensures that the mass injection rate into the disc is equal to the planet mass loss rate $\dot M_Z$. Through numerical experimentation we found that our results are insensitive to the exact injection profile as long as it is not too broad, and while $w_{\rm inj} \ll R_{\rm p}$.

Fig. \ref{fig:SS-Inject} shows the disc intensity at 1.3 mm (top panel), $\Sigma_{\rm dust}$ and maximum grain size $a$ (bottom panel) at several different times. Since initially $\Sigma_{\rm dust}$ is very low, the dust growth time is long everywhere but accelerates as more and more dust appears in the disc around the planet location. The dust grows to the (observationally required) size of $\sim 1$~mm before grain fragmentation stems grain growth. Initially, while the dust is small, dust particles diffuse both inward and outward (e.g., see the $t=1$~Myr snapshot). However, when grains become sufficiently large the radial drift starts to blow them inward from their injection site preferentially. Therefore, a quasi-steady state dust profile gets established. This profile (e.g., the red dotted curve in the bottom panel of fig. \ref{fig:SS-Inject}) is qualitatively similar to the \cite{Hogerheijde16-TWHYA} broken power-law $\Sigma_{\rm dust}$ profile (the solid green curve in the bottom panel). Likewise, the resulting disc intensity of this model at 1.3 mm reaches a steady state profile somewhat similar to the observed one. 

Fig. \ref{fig:SS-Inject-3Wavelengths} examines the disc intensity in three wavelengths (top panel) and shows the disc properties (bottom panel) at time $t=5$~Myr. Comparing the figure with Fig. \ref{fig:SS-M3-3Wavelengths}, we note that the present model has a very sharp rollover in $I_{\rm nu}$ behind the planet not because of the opacity transition at that point but because the dust surface density $\Sigma_{\rm dust}$ (black curve) nose dives at 51.5 AU. The luminosity of this model is closer to what is observed, and can also be scaled down without a significant change in the profile (except for the innermost region where the model disc is optically thick) by a simple reduction in the free parameter $\dot M_Z$. Furthermore, there is a natural casual association between the location of the planet and the rollover behind its orbit in this scenario.

The $\lambda = 9$~mm EVLA intensity of the model, on the other hand, is still problematic. Furthermore, this is a phenomenological model that contradicts physics strongly. The planet mass is kept constant at $M_{\rm p} = 3 \mearth$, whereas the dust mass actually present in the disc at $t = 5$~Myr is $M_{\rm d} \approx 25\mearth$. Increasing the planet mass at $t=0$ to a value exceeding $M_{\rm d}$ would solve the mass budget problem, but as we saw in \S \ref{sec:10ME-planet}, a planet with mass of $\sim 10 \mearth$ would produce a very deep gap in the dust disc, contradicting the observations. Further, such a massive planet would migrate inward extremely rapidly, e.g., on the time scale of $t_{\rm mig} \sim 10^5$~yrs for the disc model used in this section, invalidating the fixed orbit assumption. As the time scale for establishing the quasi-steady state dust distribution in this model is a few Myr (cf. the cyan curves in fig. \ref{fig:SS-Inject-3Wavelengths}), this is a fatal flaw -- the planet ends up in the star faster than this steady state is reached.

\section{A destroyed Core Accretion planet}\label{sec:Destroyed-CA-planet}

\subsection{Physical motivation}\label{sec:CA-motivation}

Motivated by the successes and failures of the model presented in \S \ref{sec:3ME-planet}, we now attempt to build a physically motivated model based on the Core Accretion scenario for planet formation \citep{PollackEtal96}. Core Accretion scenario planets exist in two physically very different states. After the collapse of the gas envelope around a massive solid core \citep{Mizuno80,Stevenson82,PollackEtal96}, and at the end of the runaway accretion phase, the planet mass is a few $\mj$ and its radius is only $\sim 2R_{\rm J}$.

On the other hand, before the gas accretion runaway, the planet mass is thought to be no more than $\lesssim 30-50 \mearth$, with the solids making up the majority of this mass, and the outer radius of the gas envelope $\sim$ tens of $R_{\rm J}$ \citep[e.g., see fig. 2 in][]{MordasiniEtal12a}. Previous models assumed that most of the solids get locked into the core, separating cleanly from the gaseous envelope. However, more recent work \citep{Lozovsky17-Jupiter-Z-gradients,BrouwersEtal18,Podolak19-Z-gradients} shows that most of solids are vaporised before reaching the core and are suspended as gas in the hydrogen-helium mixture. The opacity of these metal-rich gas envelopes may be significantly higher than that of the traditionally assumed Solar composition ones. Additionally, modern 3D calculations of gas and dust accretion onto cores indicate complex circulating patterns of flows which tend to recycle material from various depths in the planetary atmosphere \citep{OrmelEtal15,OrmelEtal15a,LambrechtsLega17,CimermanEtal17}. These flows make it harder for the grains to grow and sediment into the core.

Consider now a planet-planet collision energetic enough to actually unbind a Core Accretion planet. Taking a cue from the stellar collisions theory \citep{BenzHills87}, the relative velocity of the two equal mass planets at infinity, $v_\infty$, must exceed 2.3 times the escape velocity from the surface of the planet, $v_{\rm esc} = \sqrt{2 G M_{\rm p}/R_{\rm p}}$. The required collision velocity to unbind two equal mass planets is hence
\begin{equation}
    v_\infty \sim \begin{cases}  
     100 \; \text{ km/s}  & \text{for } M_{\rm p} = 1 \mj \;, \\
    3 \; \text{ km/s}   & \text{for } M_{\rm p} \sim 20 \mearth
    \end{cases}
    \label{v_infty_CA}
\end{equation}{}
The circular Keplerian velocity at 52 AU is less than 4 km/s. Collisions of CA post-collapse gas giants will lead to mergers with only a small amount of mass escaping \citep{BenzHills87}; collisions of pre-collapse planets may unbind them. It is also possible that a merger of pre-collapse planet and a massive core will lead to a common envelope like evolution, in which the cores spiral in closer together while unbinding the envelope \citep[e.g.,][]{Ivanova13-Common-Envelope}. The aforementioned high opacity makes it all the more likely that the energy deposited by the cores in the envelopes will not escape via radiation but will drive the envelope loss.

We therefore explore a model in which a pre-collapse CA planet is a source of dust. This planet may spend a long time (a few Myrs) gaining its significant mass  \citep[e.g., see][]{PollackEtal96}. In a massive disc studied in \S \ref{sec:dust-source}, such a planet would be lost into the inner disc within a very small fraction of this time due to planet migration. The migration time scale for planets scales as $\propto \Sigma^{-1} \propto M_{\rm d}^{-1}$ (eq. \ref{t1-T19}). Therefore, to make this scenario plausible we must demand the gas disc to be significantly less massive than $0.11 \msun$. We pick rather arbitrarily a value of $M_{\rm d} = 2\times 10^{-3}\msun$ while keeping the initial shape of $\Sigma$ (eq. \ref{sigma-disc0}) the same. The results do not depend very strongly on $M_{\rm d}$. Since the gas accretion rate in the disc is $\dot M_* = 3\pi \alpha c_s H \Sigma$, we must increase the disc viscosity coefficient $\alpha$ to ensure the gas accretion rate remains the same. We thus set $\alpha = 2\times 10^{-2}$. The changes to the values of the disc mass and disc viscosity parameter are important for dust dynamics. A higher $\alpha$ implies that the planet gravitational influence on the disc in its vicinity is significantly reduced. More massive planets  may be present in the disc without opening a deep gap that would contradict observations.

\subsection{Numerical results}\label{sec:CA-numerical-model}

\begin{figure}
\includegraphics[width=0.48\textwidth]{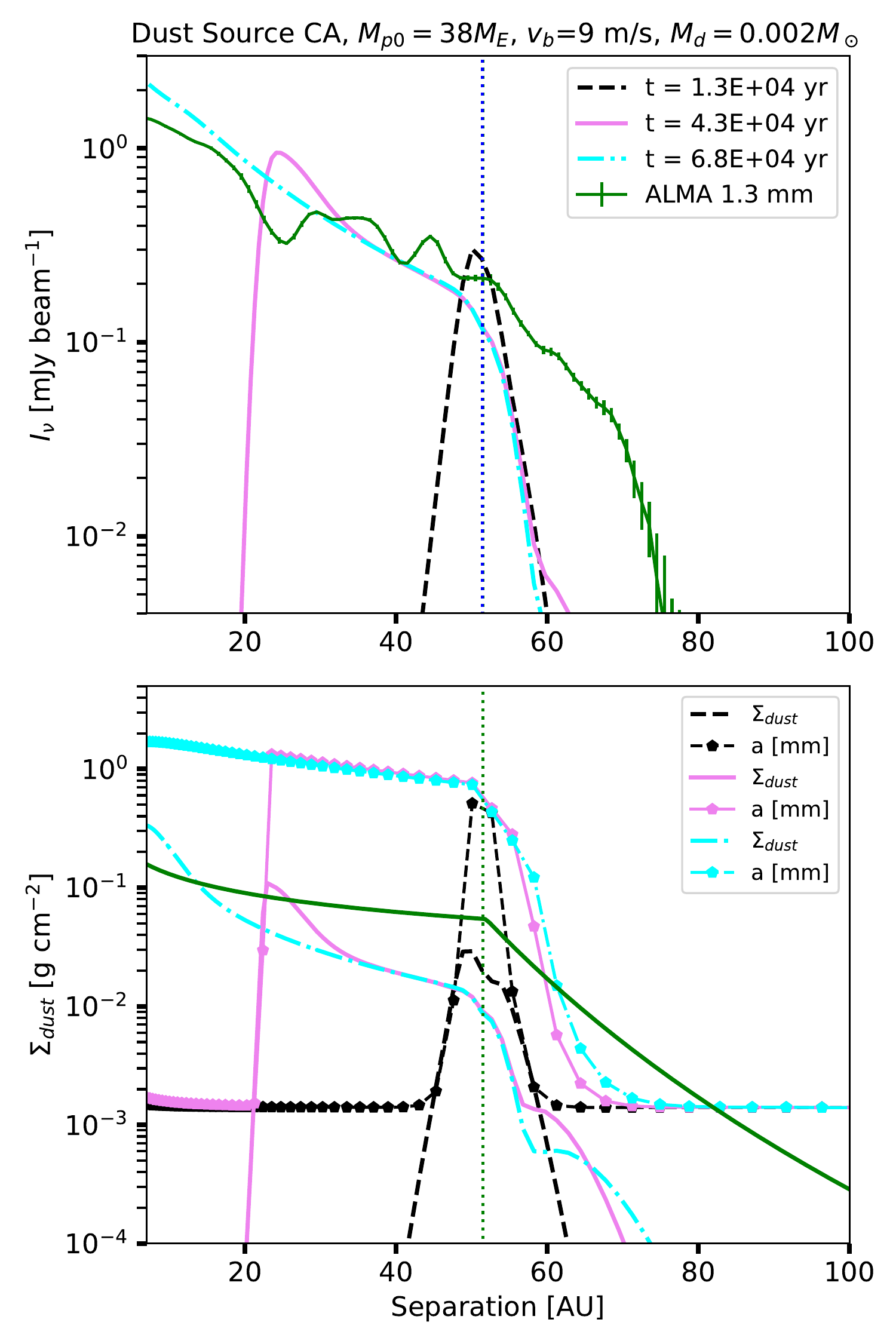}
\caption{Same as fig. \ref{fig:SS-Inject} but now for a Core Accretion Dust Source model. The planet starts off with mass of $M_{\rm p} = 38 \mearth$ and ejects both gas and dust into the surrounding disc until its mass drops to $10\mearth$.}
\label{fig:SS-CA}
\end{figure}

\begin{figure}
\includegraphics[width=0.48\textwidth]{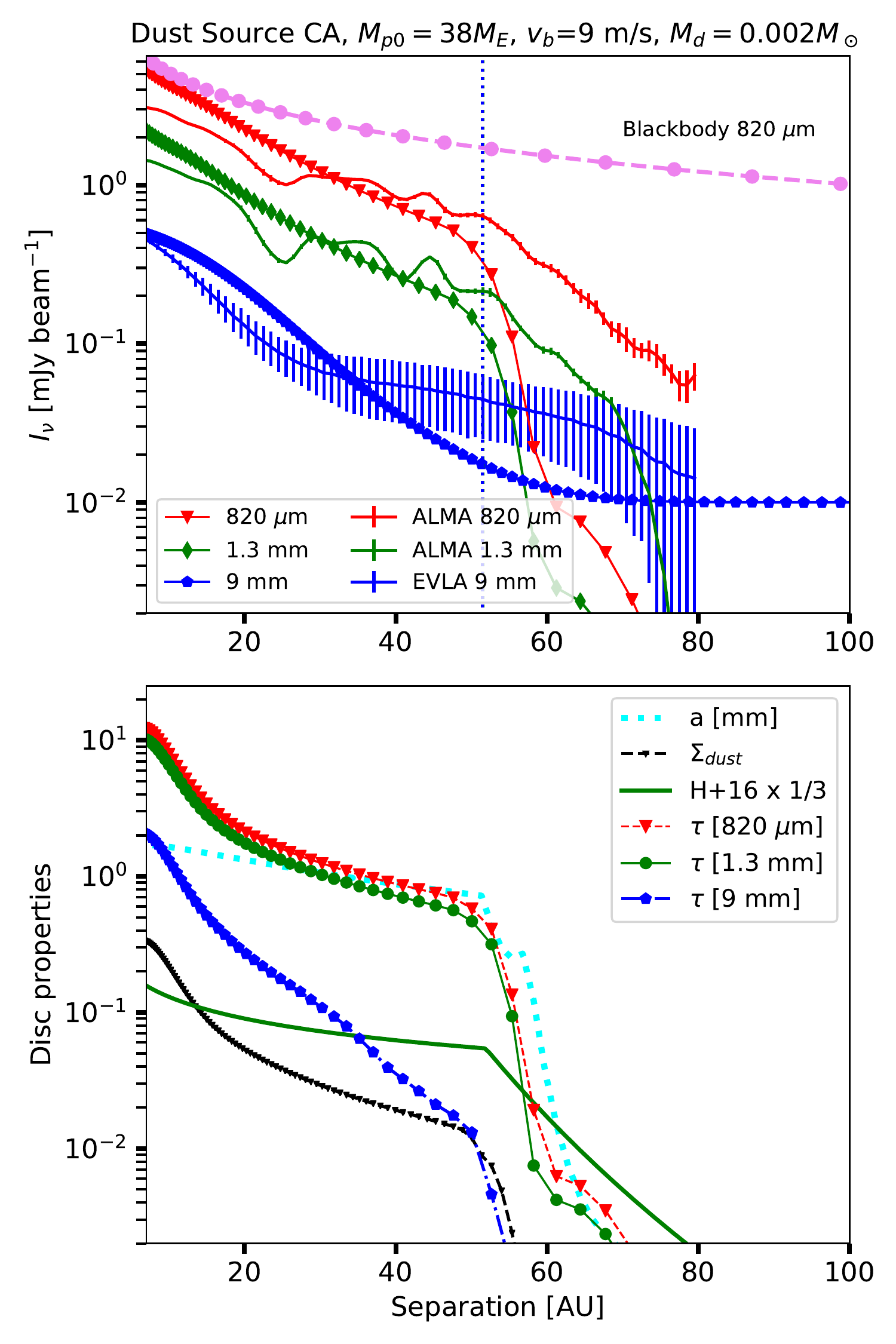}
\caption{Same as fig. \ref{fig:SS-M3-3Wavelengths} but now for the Core Accretion Dust Source model at $t=6.8 \times 10^4$~yr.}
\label{fig:SS-CA-3Wavelengths}
\end{figure}

Figs. \ref{fig:SS-CA} \& \ref{fig:SS-CA-3Wavelengths} show the results for a simulation started with initial planet mass $M_{\rm p} = 38\mearth$. For simplicity we assumed a uniform planet composition, with metallicity $Z = 0.5$. Both gas and dust are injected into the disc at a constant (and equal because $Z=0.5$) rate of $\dot M_Z = 2\times 10^{-4} \mearth$~yr$^{-1}$ until the planet mass drops to $10\mearth$\footnote{Note that in fact the metallicity $Z(M)$ of the envelope is expected to increase towards the core. Here we explore the simplest constant $Z$ case to contrast it to the more realistic scenario explored in \S \ref{sec:GI-planet-disruption}.}.  The planet mass is reduced accordingly as it loses mass. To exemplify the weak dependence of our results on the exact dust injection profile, the width of the Gaussian is here set to $w_{\rm inj} = 0.05 R_p$ (cf. eq. \ref{S-injection0}; this is three times narrower than in \S \ref{sec:dust-source}). Unlike the phenomenological model of \S \ref{sec:dust-source}, the planet is free to migrate, but on the account of the low disc mass it migrates very little during this simulation, from the starting radius of $R =53$~AU to 51.5 AU.


On the whole we see that the dust profile, and the resulting disc intensity in the three wavelengths, is quite similar to those obtained in the phenomenological massive disc model (figs. \ref{fig:SS-Inject} \& \ref{fig:SS-Inject-3Wavelengths}). As in the latter model, the disc emissivity has a very sharp -- in fact too sharp compared with the observation -- rollover behind the orbit of the planet. The similarity in the results despite the difference in the gas disc mass of a factor of 50 between the two models shows that there is a certain degeneracy in the model parameters, e.g., a higher value of $\alpha$ could be compensated for by a higher $v_{\rm br}$.

\section{GI planet disruption}\label{sec:GI-planet-disruption}

\subsection{Physical motivation and constraints}\label{sec:GI-motiv}

In \S \ref{sec:CA-motivation} we argued that a Core Accretion planet may lose a major fraction of its gas-dust envelope if two conditions are satisfied: (i) the envelope is in the extended, pre-collapse state; (ii) a significant energy is injected in it, e.g., via collision with another massive core. We now detail conditions under which a GI planet disruption could be relevant to TW Hydra's observations.

\subsubsection{The need for a very rapid primordial disc dissipation}\label{sec:GI-no-prior-disc}

In the Gravitational Instability \citep[GI;][]{Kuiper51b} theory for planet formation, massive and very young gaseous discs fragment at separations $\sim 50-100$~AU onto Jovian mass gas clumps \citep[e.g.,][]{Rafikov05,Rice05}. Hydrodynamical simulations show that these planets migrate inward very rapidly in massive discs \citep[e.g.,][]{VB06,BoleyEtal10,BaruteauEtal11}, perhaps explaining \citep{HumphriesEtal19} why wide-orbit separation gas giants are so rare in direct imaging surveys \citep{ViganEtal15,ChauvinEtal15}. On the other hand, planet-planet scatterings may allow some GI planets to survive on wide orbits \citep{ViganEtal17}, especially if the primordial disc is dispersed rapidly. For the early massive protoplanetary discs, the primary timescale on which its mass is lost \citep{ClarkeEtal01a} is the viscous time,
\begin{equation}
    t_{\rm visc} = \frac{1}{3\alpha\Omega_K}\frac{R^2}{H^2} = 5.4\times 10^{4}\hbox{ yr} \; \alpha_{-1}^{-1}
    \label{tvisc0}
\end{equation}
where $\alpha_{-1} = (\alpha/0.1)$ and the estimate is made at $R=51.5$~AU. Numerical simulations show that the $\alpha$ parameter due to self-gravity of the disc may reach values of order $\sim 0.1$ in early massive discs \citep{Gammie01,Rice05,Haworth20-Fragm-Boundary}, and even $\alpha \sim 0.2$ in the magnetised discs (Deng et al 2020). Additionally, disc depletion due to external photo-evaporation may be faster than previously thought \citep[e.g.,][]{HaworthClarke19}.

For the case at hand we must require that the primordial gas disc is long gone in TW hydra. This is because the migration time scale of a GI planet with mass initially exceeding $1\mj$ would be much shorter than 10 Myr. Indeed, if the planet did not open a gap and migrated in the type I regime then its migration time is less than 1 Myr even for a disc as low mass as a few $\mj$. If, on the other hand, the planet did open a wide gap and migrated in type II then the migration time scale is \citep[e.g.,][]{LodatoClarke04}
\begin{equation}
    t_{\rm mig2} \approx \frac{M_{\rm p}}{\dot M_*} \sim 10^{6} \hbox{ yr}\;,
    \label{tmig2}
\end{equation}{}
where we used $M_{\rm p}\sim 2\mj$ (this will be justified later) and $\dot M_*\sim 2\times 10^{-9} \msun$~yr$^{-1}$. Therefore, the planet would have been long lost into the star if the disc was there for the last 10 Myr. 

We emphasise the distinction with the CA problem setting discussed in \S \ref{sec:Destroyed-CA-planet} brought about by the different planet formation mechanisms. In the CA scenario the planet does not need to be born at $t\approx 0$. As is well known, in the classical CA model massive cores are most likely to be made at late times, e.g., at $t\sim 5-10$~Myr \citep{IdaLin04b,MordasiniEtal12} since the process of core growth is slow. Further, due to its lower mass the type I migration time scale is longer. Therefore, there is no reason to demand a complete disappearance of the primordial gas disc before the planet disruption commences in the CA framework.

\subsubsection{Why did the planet not collapse in 10 Myr?}\label{sec:GI-did-not-collapse}

GI planets are born extended, with their radius $r_{\rm p} \sim 1$~ a few AU, and cool: their central temperature is in hundreds of K \citep[e.g.,][]{HelledEtal08}. If dust growth inside the GI planet is neglected, then it cools, contracts, and eventually collapses dynamically when the endothermic reaction of H$_2$ molecule dissociation absorbs a vast amount of thermal energy of the planet \citep{Bodenheimer74}. The collapse terminates in formation of a planet that is $\sim $~Million times denser, with radius $R\sim 2 R_{\rm J}$ and an effective temperature of $1,000$ to $2,000$~K. This luminous post-collapse state is often called the "hot start" of gas giant planets \citep{MarleyEtal07}. Similar to the post-collapse gas giant CA planets, the post-collapse GI planets are unlikely to lose mass at $\sim 50$~AU from the star. 

Hence we must demand that the planet remains in the pre-collapse state before the onset of the mass loss. This is surprising given the age of the system. The evolution from birth to collapse (hot start) is usually thought to be very fast. This result is rooted in the pioneering work of \cite{Bodenheimer74} who found planet collapse time scales $\lesssim 0.5$~Myr for $M_{\rm p}=1\mj$, and even shorter for higher mass planets. However, the collapse time scale is sensitive to the dust opacity model used. More recent dust opacity calculations \citep[e.g.,][]{SemenovEtal03,ZhuEtal09,Woitke16-DIANA} indicate that dust opacity may be higher by up to a factor of $\sim 30$ at $T\sim 10-30$~K (the effective temperature of GI protoplanets) compared to the opacity employed in the 1980s \citep[e.g.,][see Appendix \ref{sec:App-dust-opacities}]{Pollack85-Opacity}. We find that these higher dust opacities lengthen the duration of the pre-collapse phase by a factor of $5-10$. Further, 3D simulations of GI planets immersed in protoplanetary discs show that these planets accrete pebbles very rapidly and become significantly enriched in dust \citep{BoleyDurisen10,BoleyEtal11a,HN18,Baehr19-pebble-accretion,Vorobyov-Elbakyan-19}. Putting these factors together we find that metal rich $M_{\rm p} \lesssim 2\mj$ gas giants may spend as long as 5-10 Myr in the precollapse configuration (see Appendix \ref{sec:App-planet-disruption} and fig. \ref{fig:Planet-disrupted-by-core})\footnote{This conclusion holds as long as grain growth and settling do not deplete $a \simlt 100 \mu$m population of grains. At higher grain sizes, Rosseland mean dust opacity may drop (cf. fig. \ref{fig:Compare_Kappas}). In that case higher $Z$ planets may actually cool more rapidly \citep{HB11}.}.

\subsubsection{GI planet disruption by a core}\label{sec:GI-core-disruption}

A number of authors have shown that pre-collapse GI planets may develop massive solid cores via grain growth and sedimentation \citep{Kuiper51b,McCreaWilliams65,Boss98,HS08,BoleyEtal10,Nayakshin10b,ChaNayakshin11a,Vorobyov-Elbakyan-19}. The time scales on which the core grows are a minimum of thousands of years but may be much longer, depending on convection and grain material/growth properties, such as $v_{\rm b}$ \citep{HelledEtal08}. If the core grows more massive than $\sim 10\mearth$, then the energy release during its formation can be too large for the pre-collapse planet -- its envelope expands and is eventually lost \citep{NayakshinCha12,Nayakshin16a,HumphriesNayakshin19-tmp}. This scenario for the core-initiated disruption of GI protoplanets is physically analogous to how cores of AGB stars eject their envelopes except for the energy source -- the gravitational potential energy rather than the nuclear energy of the core -- and the physical scales of the systems.

At present, there exists no stellar/planet evolution code that takes into account all the relevant physics that we wish to explore here. For example, \cite{HelledEtal08,HB11} present models of grain growth and sedimentation in pre-disruption isolated planets cooling radiatively. \cite{VazanHelled12} investigate how external irradiation affects contraction of these planets. These studies did not include the effects of the massive core energy release onto the planet, which is central for us here. On the other hand, \cite{Nayakshin15c,Nayakshin16a} include grain growth, sedimentation, core formation and the effects of the core energy feedback onto the envelope, but use a simplified follow-adiabats approach to model radiative cooling of the envelope, and a simpler equation of state than  \cite{VazanHelled12} do. Further, the opacities used by the two codes are different.

Here we shall use the code of \cite{Nayakshin16a} to understand the physical constraints on the pre-disruption planet mass, metallicity, and the mass of the core responsible for the planet disruption. These constrains will be seen to place significant limitations on the disrupted GI planet scenario (e.g., the planet mass is unlikely to exceed $\sim 2 \mj$). In appendix \ref{sec:App-planet-disruption} we compare for the first time the results of uniform planet  contraction calculations (no dust sedimentation allowed) computed with this code with that of the proper stellar evolution model of \cite{VazanHelled12} at the same \citep{Pollack85-Opacity} dust opacity. We find that the difference in the planet collapse time scale computed by the two codes is within a factor of two, which we deem sufficiently close given the much larger dust opacity uncertainties (\S \ref{sec:App-dust-opacities}).

\begin{figure}
\includegraphics[width=0.48\textwidth]{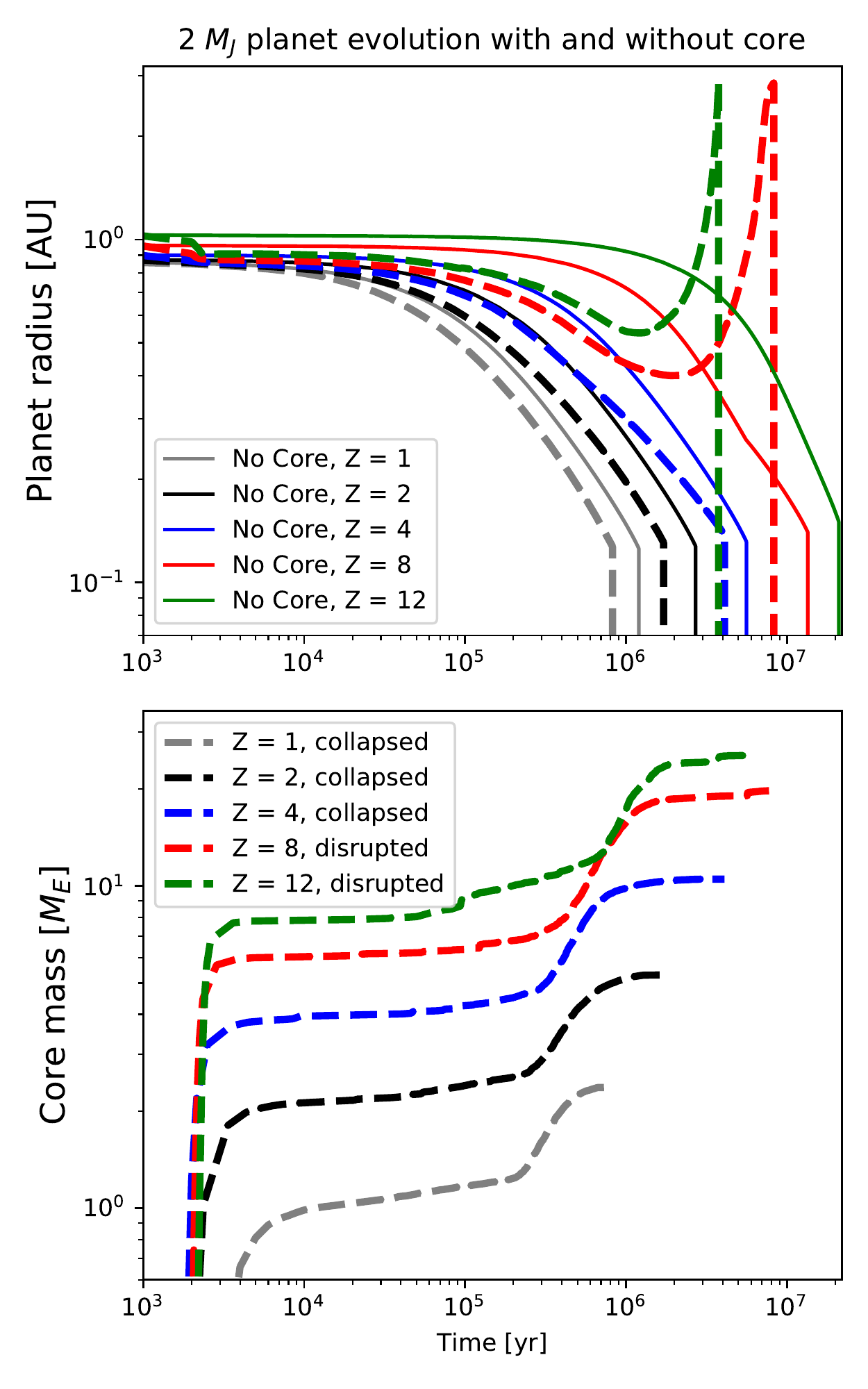}
\caption{{\bf Top:} Evolution of gas giant planet radius, $r_{\rm p}$, for different metallicities of the planet with (thick dashed) and without (thin solid) core formation. {\bf Bottom:} Core mass as a function of time for different metallicities $Z$ (in units of $Z_\odot$).}
\label{fig:Planet-disrupted-by-core}
\end{figure}

The thick dashed curves in the top panel of Fig. \ref{fig:Planet-disrupted-by-core} show the evolution of the radius $r_{\rm p}$ of a $2\mj$ planet circling the star with TW Hydra properties at 60 AU for different planet metallicities, from $Z=1$ (in units of $Z_\odot = 0.015$) to $Z=12$. All the models start with the central planet temperature of 200 K, the uniform composition and an initial grain size of $a = 0.01$~mm. The \cite{ZhuEtal09} opacity table is used for this calculation. The dust opacity is assumed to be proportional to the metallicity $Z$ of the envelope. The bottom panel of fig. \ref{fig:Planet-disrupted-by-core} presents the core mass versus time. The grain breaking velocity is here set at $v_{\rm br} =5$~m/s. The thin curves in the top panel show the same calculations but where grain growth and core formation are artificially suppressed.

Fig. \ref{fig:Planet-disrupted-by-core} shows that at Solar metallicity, $Z=Z_\odot$, the planet contracts and collapses by $t\approx 1$~Myr, whether core formation is allowed or not.  The planet evolutionary time scale increases as the metallicity of the planet increases, and so does the core mass. Formation of the core in the planet speeds up evolution of the planet in all the cases. This occurs due to a lower dust opacity in the envelope as some of the dust settles and gets locked in the core. At the lower metallicities in the figure, the core masses are relatively low, so the effect of the core formation is negligible save for the dust opacity reduction. However, for metallicities  $Z\ge 8 Z_\odot$ the core mass exceeds $20\mearth$. The gas envelope starts to expand soon after this mass is reached and is eventually unbound. For the $Z=8 Z_\odot$ model, the envelope is disrupted at about 8~Myr in this calculation. A higher metallicity planet meets its end sooner, at about 4 Myr, as its core is more massive and more luminous.

We have found that planets more massive than $\sim 2\mj$ are not likely to be disrupted by their cores. This comes about due to two factors. First of all, even at a fixed bulk composition, more massive planets contract much more rapidly, shortening the time window for the core growth \citep[this effect exists whether the core feedback is included or not, see, e.g.,][]{HelledEtal08}. Secondly, observations \citep{MillerFortney11,ThorngrenEtal15} show that more massive planets are less metal enriched than their less massive cousins. Simulations of pebble accretion onto gas giant planets \citep{HN18} also lead to the same conclusion. The dust opacity of the massive planets is hence expected to fall with planet mass, exacerbating the challenge of assembling a massive core and disrupting the planet with it.

\subsection{Deposition of matter in the secondary disc: methodology}\label{sec:GI-numerics}

There are several free parameters in this model (just like for the model in \S \ref{sec:Destroyed-CA-planet}) which we constrain by trial and error. In the beginning of the calculation, we specify the initial planet mass, $M_{\rm pi}$, and the starting position of the planet, $R_{\rm pi}$. As per \S \ref{sec:GI-core-disruption}, we use a GI planet with initial mass $M_{\rm pi} = 1.5\mj$. By experimenting we found that the disc viscosity parameter $\alpha = 2.5\times 10^{-2}$ results in gas accretion rate similar to the one observed in this system. Similarly, setting $R_{\rm pi} = 56$~AU resulted in the planet remnant being stranded at 51.5 AU; the planet initial bulk metallicity of $Z= 6 Z_\odot$ gave the right ALMA luminosity for the disc. 

We assume that at the time of disruption the planet contains a massive solid core or at least a region so metal rich that it survives the disruption of the hydrogen-rich atmosphere. We refer to the final planet mass as simply the core, and it is set to $M_{\rm c} = 0.03 \mj\approx 10 \mearth$ for definitiveness here. The part of the planet that is lost and injected into the protoplanetary disc is termed the "envelope", and its mass is $M_{\rm e} = M_{\rm pi} - M_{\rm c}$.

The mass loss rate, $\dot M_{\rm p}(t)$, is not known a priory. As for binary stars undergoing mass exchange \citep[e.g.,][]{RappaportEtal83,Ritter88}, it is a function of the planet internal structure and its orbital evolution that in itself depends on the planet-disc interaction and the mass loss rate \citep[see][]{NayakshinLodato12}. Such a fully self-consistent calculation is beyond the scope of the current paper, and we instead specify the planet mass loss rate:
\begin{equation}
\dot M_{\rm p} = 
\begin{cases}  
     - \frac{M_{\rm pi}}{t_{\rm dis}} , & \text{for } M_{\rm p} > M_{\rm c}\;, \\
    0 , & \text{otherwise } 
    \end{cases}
    \label{DotM_tot0}
\end{equation}{}
where $t_{\rm dis} = 10^5$~yrs. This mass loss rate is partitioned between that for gas (H and He) and metals (mass fraction $Z$), thus 
\begin{eqnarray}
    \dot M_{\rm g} = \left( 1-Z(M) \right) \dot M_{\rm p}\cr
    \dot M_{\rm z} = Z(M) \; \dot M_{\rm p}\;.
    \label{dotmz0}
\end{eqnarray}{}
In general we do not expect the envelope to have a uniform composition, $Z(M)$, since dust is of course able to sediment down through the gas. Hence we expect $Z(M)$ to be a function that decreases from the maximum in the core, which we simply set $Z(M_{\rm p} < M_{\rm c}) = 1$, to some minimum. For definitiveness, we choose this functional form:
\begin{equation}
    Z(M) = \left(1-Z_\infty\right) \exp\left[-\frac{M-M_{\rm c}}{M_{\rm tr}}\right] + Z_\infty\;, \; \text{ for }  M \ge M_{\rm c}\;.
    \label{fuzzy0}
\end{equation}{}
The parameter $M_{\rm tr} \ll M_{\rm e}$ describes how large the metal rich region in the centre of the planet is, and $Z_\infty$ is the metallicity at the outer reaches of the envelope where the term $\exp(-M_{\rm e}/M_{\rm tr}) \approx 0$ as we use $M_{\rm tr} \ll M_{\rm e}$. In practice, we specify the mean metallicity of the planet envelope, $\bar Z$, and $M_{\rm tr}$, from which $Z_\infty$ is computed. For the calculation presented in \S \ref{sec:GI-numerical-results}, $M_{\rm tr} = 0.03\mj$.

The dust and gas lost by the planet are deposited in a Gaussian ring around the planet location, as described by equation \ref{S-injection0}.
The normalisation constant $C$ ensures that the mass injection rate into the disc is equal to the planet mass loss rate (eq. \ref{DotM_tot0}). As per eq. \ref{dotmz0}, the injected mass is split into gas and dust. 

\subsection{Numerical results}\label{sec:GI-numerical-results}

\begin{figure}
\includegraphics[width=0.48\textwidth]{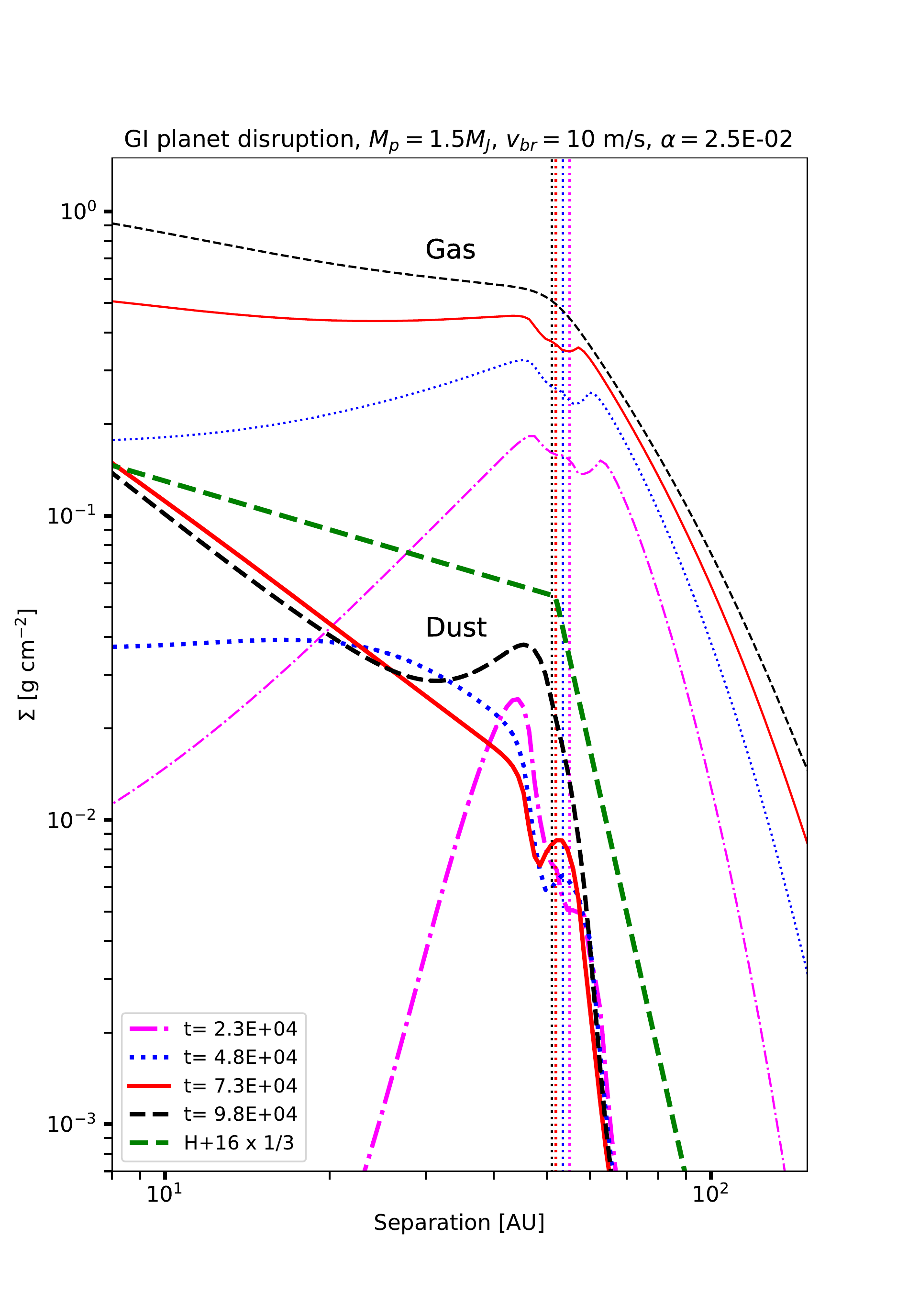}
\caption{A GI planet disruption scenario (\S \ref{sec:GI-numerical-results}). Gas (thin lines) and dust (thick lines) disc surface density profiles for selected times.}
\label{fig:TD-Sigma}
\end{figure}

Fig. \ref{fig:TD-Sigma} shows the gas and dust surface densities, $\Sigma$ and $\Sigma_{\rm dust}$, at several different times during the calculation. The vertical lines of the same type show the respective positions of the planet. 

The gas surface density (thin lines) evolution shows the dominant features of the well known viscous "spreading ring" calculation modified by the continuous mass loss from the planet. The gas spreads quickly all the way to the star and to $R\sim 200$~AU, as required by the observations of gas accretion and the large extent of the gas disc in TW Hydra (cf. \S \ref{sec:Obs}). Despite a continuous mass injection into the disc, the planet manages to make a depression in the gas surface density profile around its orbit due to gravitational torques acting from the planet onto the gas. This effect is noticeable while the planet mass is in the gas giant planet regime. By $t\approx 0.1$~Myr the planet has lost too much mass (the remnant mass $M_{\rm p} = 10\mearth$) to affect the gas surface density profile gravitationally at this relatively high value of $\alpha$ for our disc. However, the GI planet legacy lives on in the form of the significant break in gas $\Sigma$ profile; the break is coincident with the planet location. Physically, the break appears because gas flows inward towards the star inside the orbit of the planet, and outward outside the orbit.

The thicker lines show the dust surface density profile at the respective times. We see that initially the dust surface density is narrower than that of the gas, but eventually the dust spreads. This spread is mainly inward of the planet. As in \S \ref{sec:Destroyed-CA-planet}, at late times the dust dynamics is dominated by the radial drift: Once ejected by the planet, large dust particles are blown inward of the planet by the aerodynamical friction. 
The resultant dust surface density profile at $t\approx 0.1$~Myr is qualitatively similar to the broken power-law fit (the thick green dashed line) used by \cite{Hogerheijde16-TWHYA} to fit TW Hydra's ALMA dust continuum intensity profile in Band 7. The upturn in $\Sigma$ just inward of the planet is due to the assumed dust composition profile within the planet (eq. \ref{fuzzy0}) in which the dust concentration near the core is far greater than at the outer edge.

\begin{figure}
\includegraphics[width=0.48\textwidth]{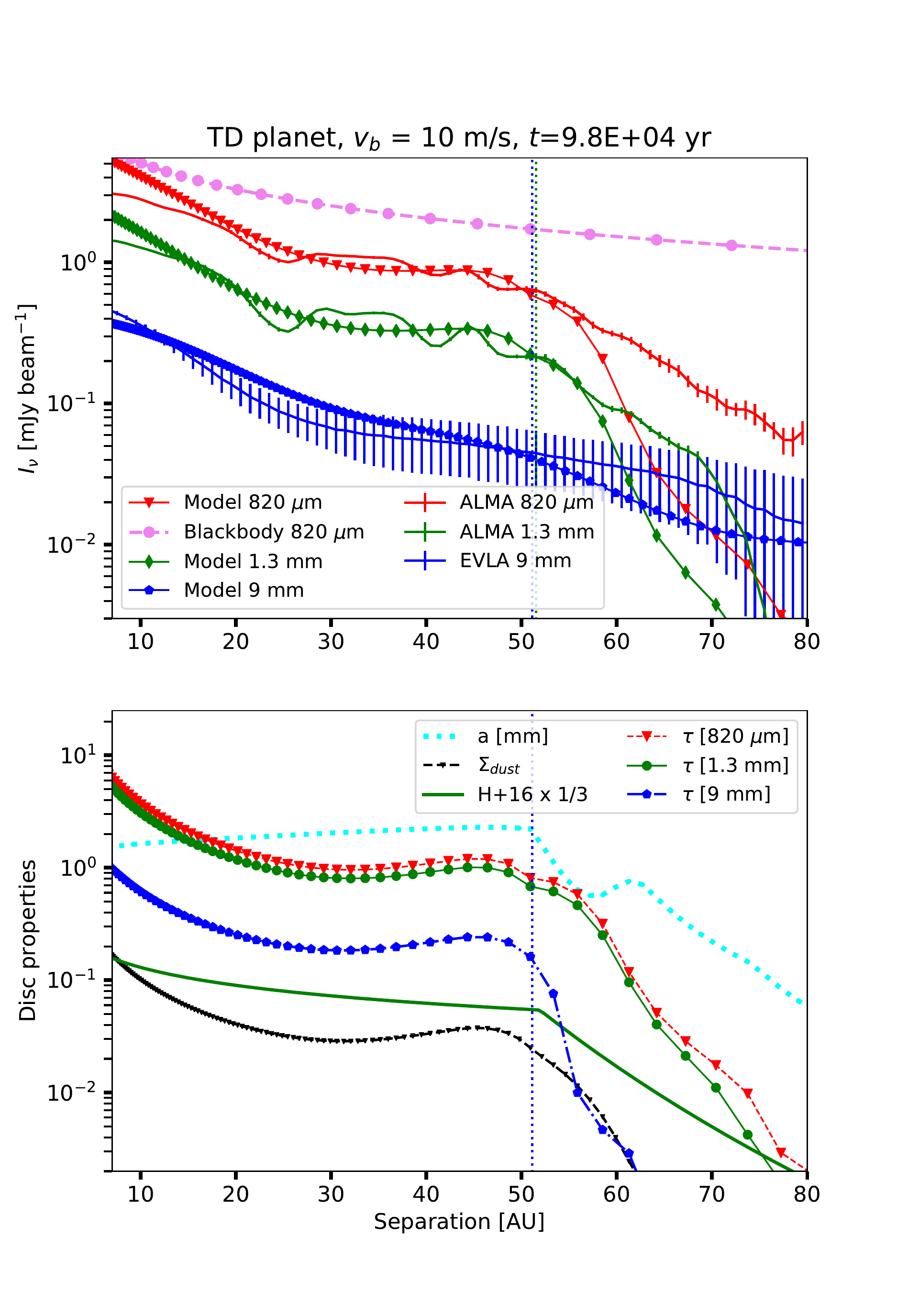}
\caption{Same as fig. \ref{fig:SS-M3-3Wavelengths} but now for the GI model (\S \ref{sec:GI-numerical-results}) at $t=6.8 \times 10^4$~yr.}
\label{fig:TD-3Wavelengths}
\end{figure}

Fig. \ref{fig:TD-3Wavelengths} shows the resultant disc emissivity profile in the three wavelengths in the top panel and the disc properties in the bottom panel. The model fits the data reasonably well except for the rollover, which is too sharp, just like the model in \S \ref{sec:CA-numerical-model}. This model is optically thin, as is observationally desired. The total mass of the dust in the disc in this model is about $15\mearth$, much smaller than $\sim 80\mearth$ estimated by \cite{AndrewsEtal16} and also smaller than $\sim 35\mearth$ estimated by \cite{WoitkeEtal19}. This is mainly due to the different dust models used in these studies. The standard DIANA dust opacity \citep{Woitke16-DIANA} are higher by a factor of a few than that used by \cite{AndrewsEtal16,Hogerheijde16-TWHYA}. Although we use the public DIANA opacity code to compute our dust opacities here, we use the standard $a^{-3.5}$ dust size distribution whereas \cite{WoitkeEtal19} found that $a^{-4}$ power law was a better fit in their modelling.

\section{The ALMA image of T19 feature}\label{sec:dust-near-T19}

Here we discuss the implications of the 2D morphology of the excess emission observed by \cite{TsukagoshiEtal19}. We use these as additional probes of the scenarios explored in this paper.

\subsection{The disc is optically thin}\label{sec:disk-is-opt-thin}

As found by \cite{TsukagoshiEtal19}, the excess  has a radial half width of $\sim 0.5$~AU and an azimuthal half-width of $\sim 2.2$~AU. The disc pressure scale height for TW Hydra is $H \sim 3.5$~AU at separation of 51.5 AU, and thus the observed feature is significantly smaller than $H$. We note that this immediately implies that the disc (but not necessarily the feature) is optically thin at 51.5 AU. This is because photons emitted from the disc midplane perform a random walk in an optically thick disc until they escape vertically out of the disc. Therefore, an image of a point source placed in the midplane of such a disc would be broadened by $\sim H$ at least. This forms an independent confirmation, in addition to the arguments spelled out in \S \ref{sec:3ME-planet} that TW Hydra's disc is optically thin and thus the rollover in $\Sigma_{\rm dust}$ observed behind the T19 feature could not be due to the disc becoming optically thin at this radius.


\subsection{A vortex or a circum-planetary disc?}\label{sec:not-circum-planetary-disc}

Vortices \citep[e.g.,][]{Li01-RosbyWave-Vortices} 
have been suggested to trap dust material in the protoplanetary discs and thus produce bright excess in the dust continuum emission \citep{BaruteauZhu16-vortices}. Furthermore, vortices are azimuthally elongated structures, with axis ratio $\sim 4-6$ \citep{Richard13-vortices}, exactly as observed. However, the radial half-width of vortices is at least $H$, and likely $\sim$ twice that \citep[e.g., fig. 3 in][]{Lin-Min-Kai-12-vortices}. For the same reason a vortex would also look much more extended in the azimuthal direction \citep[see fig. 4 in][]{BaruteauZhu16-vortices} than observed. Just like with a gap edge created by a planet, we expect a hole in the dust density distribution inward of the vortex \citep[fig. 10 in][]{BaruteauZhu16-vortices}, which is not observed.

Circum-planetary discs are believed to be at most $\sim 1/3$ \citep{Bate03,AyliffeBate09}, and more likely $1/10$ of the planet Hills radius, as shown by the more recent higher resolution calculations \citep{Wang14-circum-disc,OrmelEtal15}. Thus, the planet would have to exceed the mass of $\sim 1.5\mj$ to account for just the radial size of the feature. This does not account for the much larger azimuthal extent of the T19 excess emission. Such a high planet mass would produce a noticeable gap at the disc gas surface density and the dust intensity profile near the planet even for disc viscosity as high as $\alpha = 0.025$. This is not observed. Finally, \cite{TsukagoshiEtal19} also points out that the total flux from the circum-planetary disc is insufficient to account for the total flux in the feature.

\subsection{The dust trail of a planet losing mass}\label{sec:dust-trail}

Here we consider the dynamics of grains lost by a low mass planet on a circular orbit embedded in a laminar gas disc around it. The most likely physical origin for a dust particle outflow from a planet is a thermally driven gas outflow that picks and carries the dust with it. The dynamics of grains in the planet vicinity, at radii between the planet radius and the Hills radius, $r_{\rm p} \lesssim r \ll R_{\rm H}$, clearly deserves a separate detailed investigation which is outside the scope of this paper. Here we are concerned with how the flow may manifest itself to ALMA on scales of much larger than $r_{\rm p}$. We perform a 2D calculation of dust particle orbits assuming their trajectories lie in the midplane (note that $R_H \ll H$ for a low mass planet).

The dust particle size at the outflow is likely to be much smaller than the mm-sized particles that ALMA sees in TW Hydra's disc. This is because the inner region of the planet is expected to be sufficiently hot to vaporise even the most refractory dust \citep{BrouwersEtal18}. This is relevant because for both the CA and GI planet losing mass scenarios the time when the model fits the data best is close to the end of the mass loss phase from the planet, when the most central regions of the planet are lost into the disc. 

As the outflow leaves the planet, the gas density drops, and so does its optical depth. Due to adiabatic expansion and radiation (the outflow eventually becomes transparent to radiation) the gas temperature drops with distance from the planet and the metals re-condense into dust particles. The grains then grow rapidly. In the disc geometry, the grain growth time scale is $t_{\rm grow} \sim (1/\Omega_K) (\Sigma/\Sigma_{\rm dust})$ \citep{BirnstielEtal12}. For the gas just lost by the planet the gas-to-dust ratio is not very large as the planet central regions are very metal rich in our model, therefore $t_{\rm grow}$ may be expected to be of order a few orbital times in the disc, $2\pi/\Omega_K$.


The dynamics of dust particles is most sensitive to the Stokes number, $\St$, and hence we reformulate the problem in its terms.
The dust particles are ejected by the planet with initial Stokes number $\St_{\rm min} = 10^{-3}$, and grow to a maximum size corresponding to the  maximum Stokes number of $\St_{\rm max} = 0.1$. We describe the particle growth process as a time-dependent $\St$ number
\begin{equation}
    \ln \St = (1-q') \ln \St_{\rm min} + q' \ln \St_{\rm max}\;,
    \label{St-vs-t}
\end{equation}{}
where $q' = t'/(t_{\rm grow} + t')$, where time $t'$ counted from the time the grain was ejected by the planet, $t_{\rm grow} = N_{\rm g} (2\pi/\Omega)$ is the growth time scale, with $N_{\rm g} =4$. For reference, $\St = \sqrt{\St_{\rm min} \St_{\rm max}} = 0.01$ at $t' = t_{\rm grow}$.

\begin{figure*}
\includegraphics[width=0.98\textwidth]{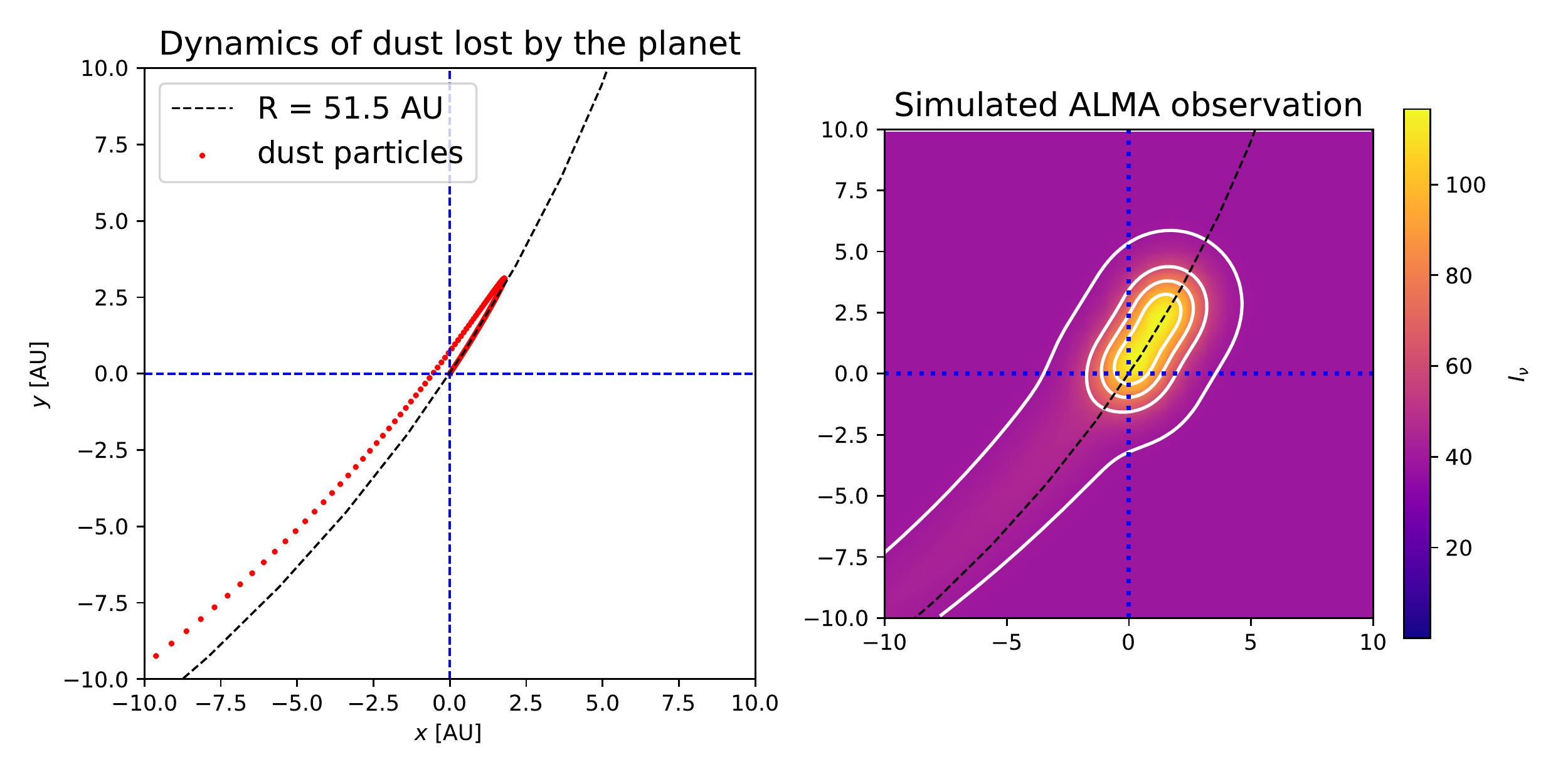}
\caption{2D dust particle dynamics in the vicinity of the planet as seen face-on to the disc. The disc and the planet rotate in the clock-wise direction. {\bf Left:} Locations of individual dust particles emitted by the planet at a steady rate are shown with red dots. The planet is located at the centre of the coordinates in the image and travels along the circular black-dashed path. {\bf Right:} The simulated ALMA image of the emission from the dust distribution shown in the left panel. The image bears a certain resemblance to observations shown in the right top panel of fig. \ref{fig:Obs}.}
\label{fig:Zoom-in}
\end{figure*}

The planet is assumed to be of a sufficiently small mass that we can neglect its dynamical influence on the surrounding gas. Similarly, we neglect the planet physical size compared with the scales of interest, assuming that dust is emitted from a point (planet position). The planet is on a circular orbit around the star at $R=51.5$~AU. We integrate the standard 2D equations of motion for individual dust grains lost by the planet. Grains are ejected by the planet at a steady rate. As expected, the grains are blown inward by the radial drift and eventually disappear into the star but here we are interested in the dust particle morphology in the immediate vicinity of the planet to compare to the T19 ALMA image of the excess emission region.

The left panel of Fig. \ref{fig:Zoom-in} shows the location of the dust particles integrated as described above. The coordinates are centred on the planet which is on a prograde circular orbit that follows the dashed circular path. We observe that the dust particles perform a U-turn as seen from the planet location. Since initially the dust particles are small ($\St \ll 1$), they are very tightly coupled to gas at $t\ll t_{\rm grow}$. In the frame of the planet they start to lag behind the planet because the dust is picked up by the gas in the disc and so travels with the velocity
\begin{equation}
    v_{\rm gas} = v_{\rm K} \left(1 - \eta \frac{H^2}{R^2}\right)^{1/2} < v_{\rm K},
\end{equation}{}
which is smaller than the planet orbital speed $v_{\rm K}$. However, as the dust particles grow, they start to drift inward of the planetary orbit. When the dust particle drifts to radius $R' < R$ such that its angular speed there exceeds that of the planet, that is, $v_\phi(R')/R' > v_{\rm K}(R)/R = \Omega_{\rm K}(R)$, it starts to orbit around the star faster than the planet. Hence the particle overtakes the planet eventually. 

This orbital motion of the dust looks like a tight U-turn around the planet. Additionally, as the particle sizes grow as they move farther and farther away from the planet, the speed differential between the planet and the dust particles increase. This leads to dust particles spending more time in the U-turn region than in the region in front of the planet.The dust density is hence larger in close proximity to the planet and decreases with distance along the dust tail.

The right panel of fig. \ref{fig:Zoom-in} shows a simulated ALMA image of the dust dynamics smoothed by the ALMA beam for TW Hydra. We added a point source to the planet location and a uniform background. We see that the U-turn of the dust produces an emission feature elongated in the azimuthal direction, somewhat analogous to the observations shown in fig. \ref{fig:Obs}. The exact brightness of the extended tail compared to the U-turn region, the pitch angle of the tail with respect to the azimuthal direction, and the length of the U-turn region do depend on the parameters on the dust growth model used. However, a good qualitative match to the shape of the \cite{TsukagoshiEtal19} excess emission is obtained for a wide range of  model parameters, not requiring fine tuning. The weak excess emission in the dust tail running in front of the planet and pointing inward of its orbit is a unique testable prediction of our model.

\section{Discussion}\label{sec:Discussion}

TW Hydra hosts a protoplanetary disc resolved by ALMA with a record breaking resolution of about 2 AU. The disc has a cliff-like rollover beyond about 50 AU and a significant excess emission resolved by ALMA into a blob with sizes $\sim 1 $AU radially by $\sim 4$~AU azimuthally. The excess is suspected to be a young planet and is located at 51.5 AU, right at the edge of the dust disc. The protostar continues to accrete gas at a respectable rate of $\sim 2\times 10^{-9}\msun$~yr$^{-1}$ despite being one of the oldest protstars known ($\sim 10$~Myr old). Furthermore, its dust disc is nearly two orders of magnitude more massive than the median for  Class II sources \citep{WilliamsEtal19-ODISEA}, most of which are younger than TW Hydra. Here we have shown that we can use these known and unique properties of the system to constrain scenarios of the protoplanetary disc evolution.

\subsection{The quasi Steady State scenario}\label{sec:SS-discussion}

In this scenario (\S \ref{sec:steady}) the protoplanetary disc in TW Hydra is primordial, e.g., $\sim 10$~Myr old. Previous work \citep[e.g.,][]{Powell19-DustLanes} and analytical arguments on the presence of $\sim$ mm-sized dust in the disc (\S \ref{sec:d-drift}), and the observed gas accretion rate (\S \ref{sec:gas_accretion}), require the gas disc mass $M_{\rm d}$ to exceed $\sim 0.1\msun$ in this case. We found this scenario to be challenged by the data and other results in the field:

\begin{enumerate}
\item\label{item:wrong-spectr} {\bf The observed cliff-like rollover} in the dust continuum emission beyond $\sim 50$~AU is very puzzling for such an old disc that is known to extend to $\sim 200$~AU in CO and other molecular tracers. Numerical experiments (\S \ref{sec:3ME-planet}) and analytical arguments (\S \ref{sec:SS-the-drop}) show that one expects a power-law like decline in $\Sigma_{\rm dust}$ with radius in this case. We found that the scenario may produce a sharp break in the disc emissivity if the disc becomes optically thick inside 50 AU (fig. \ref{fig:SS-M3-3Wavelengths}). However, for TW Hydra this over-predicts the total flux from the disc by a factor of $\sim 3$, contradicts earlier conclusions from photometry data (\S \ref{sec:3ME-planet}), and the relatively small size of the \cite{TsukagoshiEtal19} feature (\S \ref{sec:disk-is-opt-thin}). All of these observations require an optically thin disc in the ALMA bands. Further, this model disc becomes optically thin at longer wavelengths and hence predicts a steep decline in  the disc intensity with radius at $\sim 50$~AU, whereas the 9 mm EVLA data show a very gradual decline in that region (blue curves in fig. \ref{fig:SS-M3-3Wavelengths}).

\item\label{item:planets} {\bf The presence of planets.} \cite{MentiplayEtal18} inferred the masses of $M_{\rm p} \sim 4\mearth$ for the two putative planets located inside the observed gaps at 24 and 41 AU.  \cite{TsukagoshiEtal19} estimated the planet mass at 51.5 AU to be $M_{\rm p} \sim 1$ Neptune masses. The type I migration time scale for these planets is $\sim 1$ to a few\% of TW Hydra age (\S \ref{sec:typeI-migr}). To observe even one of these three putative planets, we need to be quite lucky. To observe all three of these planets, we need to postulate that they were all born essentially simultaneously. This is unlikely because the rates of planet embryo assembly are strongly separation-dependent \citep{IdaLin04a,MordasiniEtal12a,LambrechtsEtal14,NduguEtal19}.

\item\label{item:T19-and-rollover} {\bf The association of the \cite{TsukagoshiEtal19} feature with the dust disc rollover.}  Planets are expected to block the inward flow of dust, producing gaps at the location of the planets, and bright rings just beyond the gaps \citep{RiceEtal06,PinillaEtal12,DSHARP-6,Dsharp7}. Our numerical experiments in \S \ref{sec:10ME-planet} confirmed these well known results. These dust emission characteristics are not observed in TW Hydra, where the emission plunges instead of rising beyond the planet location.

\item\label{item:T19-shape} {\bf The size and shape of the extended excess emission detected by \cite{TsukagoshiEtal19}.} The half-sizes of the feature at 1.3 mm continuum is $\sim 2.2$~AU in the azimuthal direction and $\sim 0.5$~AU in the radial emission. This extent is too small for a vortex but too large for a circum-planetary disc of a Neptune mass planet (\S \ref{sec:not-circum-planetary-disc}). Increasing the planet mass above $1\mj$ may result in the disc large enough but its elongation along the azimuthal direction by a factor of 4 contradicts numerical simulations of circum-planetary discs which show no such elongation \citep{AyliffeBate09,Wang14-circum-disc,OrmelEtal15a,Szulagyi14}. Additionally, such a high mass planet is ruled out based on the protoplanetary dust disc morphology as explained in \ref{item:T19-and-rollover}.

\item\label{item:too-low-alpha} {\bf A very small disc viscosity.} The observed gas accretion rate onto the star is surprisingly low if the disc mass is really as high as $0.1\msun$, and requires the disc viscosity parameter $\alpha \lesssim 10^{-4}$ (see \S \ref{sec:d-drift}). The results of modelling ALMA observations of other bright discs with annular gaps and rings show that for particle sizes $a\simgt 2$~mm the disc viscosity parameter must be larger than $\sim 10^{-3}$ \citep[fig. 7 in][]{DSHARP-6}.

\item\label{item:very-low-vbreak} {\bf Surprisingly small grains.} Analytical arguments (\S \ref{sec:d-size}) and numerical models show that in a disc as massive as $0.1\msun$ the grains should grow rapidly to sizes much larger than $\sim $ a few mm, and be lost into the star too soon, unless the dust fragmentation velocity is $\sim 0.3$~m/s. The latter value is significantly lower than the values obtained in laboratory experiments \citep[e.g.,][]{BlumWurm08} and $\sim 10$~m/s typically used in the protoplanetary disc literature \citep[e.g.,][]{BirnstielEtal12,DrazkowskaEtal14,Rosotti19-Opacity-Cliff}.

\item\label{item:no-spirals} {\bf Dusty rings rather than spirals.} \cite{Veronesi19-light-discs} show for TW Hydra and other ALMA discs that gas discs can be "weighted" by understanding the response of the mm-sized grains to the planets embedded in these discs. They find that in massive discs the mm-sized grains would tend to be in spiral features driven by the planets, whereas in low mass gas discs they would conform to the shape of rings. Based on the absence of spirals and presence of rings the authors conclude that the disc masses are $\sim 1\mj$.

\item\label{item:oddity} {\bf Oddity compared to other discs.} The dust mass of TW Hydra is extraordinarily large. Using the pre-DIANA opacity model, this mass is estimated at  $\sim 80 \mearth$ \citep{AndrewsEtal12-TW-Hya,AndrewsEtal16}. The mean dust mass of class 0 sources was recently estimated at $26\mearth$ \citep{Tobin20-Class0-masses}; these sources are a factor of $\sim 100$ younger than TW Hydra. The more comparable yet still younger by a factor of several class II discs have dust disc masses almost 2 orders of magnitude lower than that of TW Hydra \citep{WilliamsEtal19-ODISEA}. Using DIANA \citep{Woitke16-DIANA} opacities we find a factor of $\sim 6$ lower dust mass for TW Hydra; however all the other  results cited above should then be scaled down as well, leaving TW Hydra's dust mass excess just as large.

\end{enumerate}

\subsection{The phenomenological planet losing dust scenario}\label{sec:disc-phen-planet-losing-discussion}

In \S \ref{sec:dust-source} we used the same massive disc scenario as discussed in \S \ref{sec:SS-discussion}, but assumed that the protoplanetary disc is dust-free before a source of dust of unspecified nature starts ejecting dust. This produced a better match to the data, resolving qualitatively the problems listed in \ref{item:wrong-spectr}. Large dust particles drift inward, naturally explaining why the T19 feature is positioned at the dust disc rollover, alleviating \ref{item:T19-and-rollover}. However, the model violates mass conservation and second Newton's law, and does not resolves the other challenges from \S \ref{sec:SS-discussion}.

\subsection{A Core Accretion planet losing dust}\label{sec:CA-losing-dust-discussion}

We argued in \S \ref{sec:CA-motivation} that a massive core can lose its pre-collapse massive dusty gas envelope if a catastrophic release of energy occurs in its core, due to e.g., a merger of the core with another massive core. We dropped the assumption of a massive gas disc, which becomes unnecessary if the dust in the protoplanetary disc of TW Hydra is of a recent rather than primordial origin. In the particular example of the numerical calculation in \S \ref{sec:CA-numerical-model} we considered a pre-collapse planet of the total initial mass of $38\mearth$ to lose its half dust/half gas envelope (planet metallicity $Z=0.5$) until its mass dropped to $10\mearth$, at which point the mass loss was turned off. The initial disc mass was set at $0.002 \msun$, which required disc viscosity of $\alpha = 0.02$ to yield the correct gas accretion rate onto the star. 

This vastly improved the results, resolving all issues \ref{item:wrong-spectr}-\ref{item:oddity}, e.g., producing a reasonable match to the observed spectra with a now reasonable value for the grain fragmentation (breaking) velocity, $v_{\rm br} =10$~m/s, a value of $\alpha$ in accord to the constraints from DSHARP modelling \citep{DSHARP-6}. Due to the much lower gas disc mass, the planet migration time is comparable to the age of the system, not requiring a miracle of several planets being born at the same time. The Stokes number of mm-sized grains satisfied the \cite{Veronesi19-light-discs} constraint. A disruption of a massive pre-collapse planet via a catastrophic collision with another planet is not likely to be a common outcome for the Core Accretion scenario, and this may explain why TW Hydra is such an oddity \ref{item:oddity}. Further, recently \cite{Demidova19-CA-collision} showed that catastrophic collision of planetary embryos in a protoplanetary discs releases enough dust to be observable with ALMA.

\subsection{A Gravitational Instability planet disruption}\label{sec:GI-planet-discussion}

In \S \ref{sec:GI-planet-disruption} we considered a disruption of a gas giant planet formed by the GI scenario. Since GI planets are presumably born in very young, class 0/I discs, this planet would have survived at such a wide orbit only if the disc was dissipated very rapidly in this system (\S \ref{sec:GI-no-prior-disc}). This therefore requires that TW Hydra had no protoplanetary disc before the planet disruption. It is also possible that there were more GI planets early on, and that one of them was scattered on a wider orbit than the rest, boosting its chances of survival far out.

Just as with the Core Accretion planets, the {\em pre-collapse} GI planets are extended and are susceptible to their envelopes being destabilised if enough energy is injected into the planet centres. Massive solid cores ($M_{\rm core}\gtrsim 10\mearth$) were previously shown to be capable of disrupting the planet envelopes \citep[e.g.,][]{NayakshinCha12,Nayakshin16a,HumphriesNayakshin19-tmp}. For a very old system such as TW Hydra, we found that only very metal rich $(Z\sim 0.1$) GI planets with masses no larger than $2\mj$ can be disrupted via this mechanism at $t\sim 10$~Myr (\S \ref{sec:GI-did-not-collapse} and \S \ref{sec:GI-core-disruption}). 

In \S \ref{sec:GI-numerical-results} we found that disruption of a planet with an initial mass $M_{\rm p} = 1.5\mj$, initially orbiting TW Hydra at 56 AU resulted in a gas disc quickly spreading both inward, to fuel gas accretion onto the star at rates close to those observed in the system, and outward to $\sim 200$~AU. As with the CA planet losing dust scenario, the dust lost by the planet grows to mm sizes and then streams only inward of the planet due to the aerodynamical friction with the gas. 

Although the physics of the models differs, the two scenarios give equally promising explanations for the observed spectra of the source and yield attractive explanations to all the points \ref{item:wrong-spectr} -- \ref{item:oddity} raised as difficulties of the standard quasi-steady state scenario.

\subsection{The dust morphology of the T19 excess emission source}\label{sec:T19-source-discussion}

Finally, we investigated the dynamics of dust grains lost by a low mass planet in 2D in \S \ref{sec:dust-near-T19}. We argued that dust particles must be carried away from the planet by a gas outflow, and must therefore be microscopic initially. We then argued that when released into the disc the dust will grow to larger sizes as constrained by the disc properties. The dust particles were found to perform a U-turn around the planet, first being dragged along by the gas flowing past the planet, but then overtaking the planet a little inward of its orbit when they have grown sufficiently to drift through the gas. When convolved with the ALMA beam at 1.3 mm this results in an emission excess elongated along the orbit and predicts a weak tail extending in front and a little inward of the planetary orbit.

\subsection{Planet-losing mass scenario disadvantages, uncertainties and future tests}\label{sec:disadvantages}

While we found a number of compelling spectral and physical arguments to favour the planet disruption scenarios over the traditional quasi steady-state framework for disc evolution with planets that only gain mass, there are many issues that nee further investigation. First of all, more effort needs to be invested in detailing the conditions under which Core Accretion and Gravitational Instability planets can be disrupted with such a significant amount of dust lost as well. For this to be the case, the dust in the envelope  must be well coupled to the gas or else it settles into the core, and the envelope would also cool too rapidly and hence collapse. This tight dust-gas coupling is probably natural in the envelope centre where dust may sublimate and be in the gas phase anyway, but in the outer regions of the envelope the dust must remain sufficiently small.

One should also aim to constrain the mass loss rate and parameters such as the outflow speed from first principles. Our 2D dust dynamics calculation in \S \ref{sec:dust-trail} assumed that gas dynamics near the planet is dominated by the laminar shear flow of gas around the protostar. It is quite possible that 3D hydrodynamical calculations of a gas-dust outflow from a planet in the disc may lead to spiral features, which may be very different in nature to those launched by the planet's gravity. Recent detection of spiral density features in TW Hya disc by \cite{Teague19-TW-Hya-spirals} may test this scenario.

Both CA and GI planet disruption scenarios for TW Hya hinge significantly on the dust opacity model used here \citep{Woitke16-DIANA}, which predicts opacity larger than much of previous work. \cite{Woitke16-DIANA} show that their "standard" opacity is larger than that used by \cite{AndrewsWilliams05} at $850 \mu$m by a factor of 1.6. We further found that to match the normalisation of the disc intensity profiles in the three wavelengths simultaneously, an amorphous carbon fraction of 26\% is required, which is very close to the 25\% fraction found by \cite{WoitkeEtal19} for TW Hya. This yields a further increase by a factor of $\sim 2-3$ in the dust absorption extinction in $\sim$ mm wavelengths (e.g., see fig. 3, the green curve, in \cite{Woitke16-DIANA}). As a result, the dust disc mass of TW Hya in our models is only $\sim 15\mearth$, low enough to be accounted for by dust rich envelopes of massive planets.

This contrasts strongly with the results of \cite{Ueda20-TW-Hya} who have recently obtained a dust mass of $\sim 150\mearth$ just inside the inner 10 AU of TW Hya. Extended to the outer dust disc edge, this estimate is two orders of magnitude larger than the values obtained here. \cite{Ueda20-TW-Hya} emphasize the inclusion of dust scattering as the main driver of their much higher dust mass compared with previous literature. However, these effects are also included here via the \cite{Zhu19-scattering-albedo} formalism. In fact, all of our reasonably successful fits to TW Hya disc intensity profiles are becoming optically thick inward of 10 AU (cf. figs. \ref{fig:SS-Inject-3Wavelengths}, \ref{fig:SS-CA-3Wavelengths}, \ref{fig:TD-3Wavelengths}), in close agreement with \cite{Ueda20-TW-Hya}. The \cite{WoitkeEtal19} calculations also include dust scattering, and their dust mass for TW Hya is similar to ours. Therefore, the more likely source of the disagreement is in the dust opacity model. As an example, \cite{Ueda20-TW-Hya} dust absorption opacity is 65 times lower than ours at the wavelength of 3.1 mm. These differences show that constraining the actual dust opacity in TW Hya would go a long way towards testing our model; if dust extinction opacity is significantly lower than used here then the dust disc mass budget is simply too large to originate from a disrupted planet of any sort.

\begin{figure}
\includegraphics[width=0.5\textwidth]{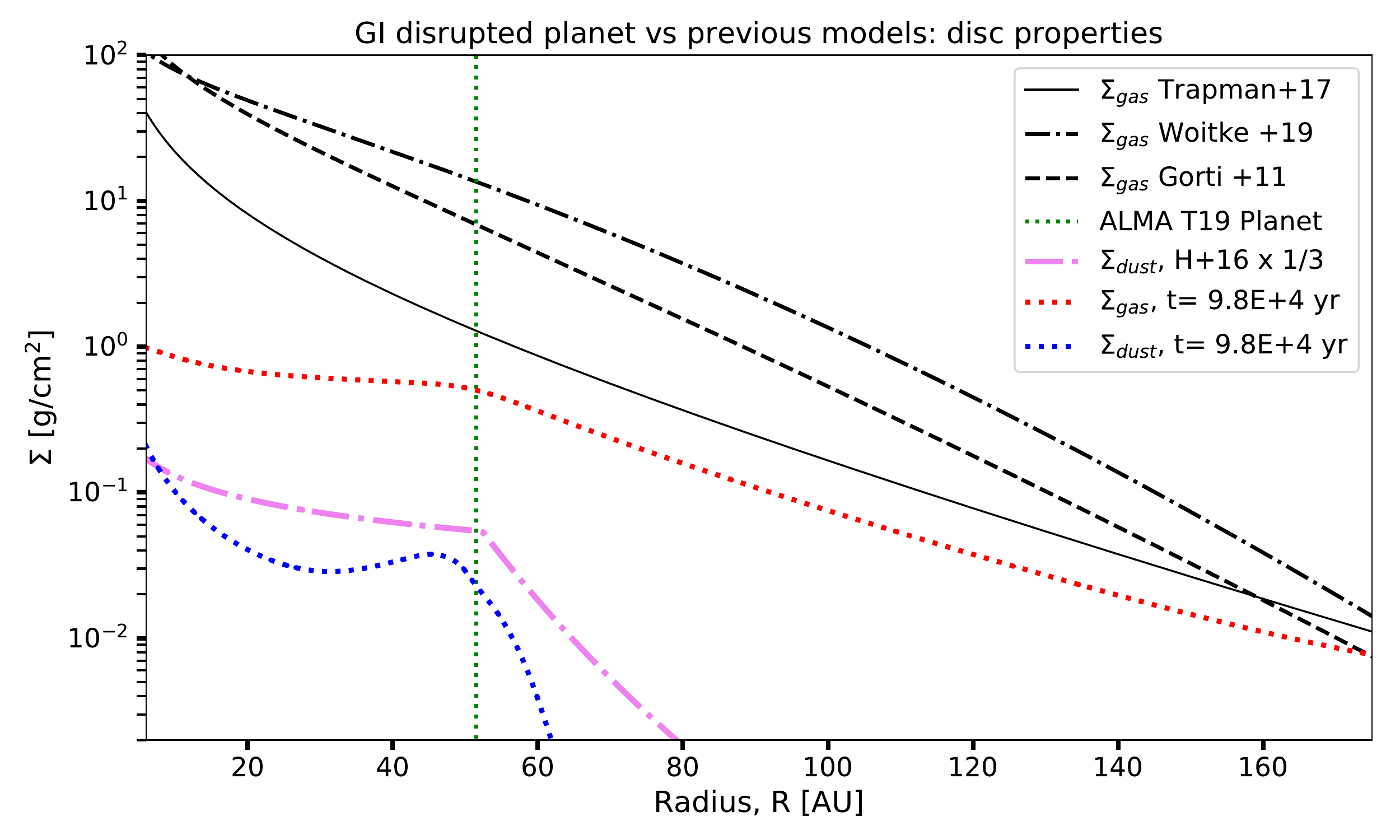}
\caption{Comparison of the gas (red dotted) and dust (blue dotted) disc surface density profiles of the GI planet disruption model (fig. \ref{fig:TD-3Wavelengths}, \S \ref{sec:GI-planet-disruption}) with that of previous authors for TW Hya. While our model matches the dust density profile from Hogerheijde et al (2016) reasonably closely, our gas surface densities are lower by a factor of a few than Trapman et al (2017).}
\label{fig:Model-vs-Obs}
\end{figure}

Further, a detailed chemo-dynamical modelling of TW Hydra in the context of a disrupted planet scenario is needed to ascertain that it may explain the extremely rich data set for this well observed source \citep[e.g.,][]{AndrewsEtal12-TW-Hya,BerginEtal13,Menu14-TWHya,WoitkeEtal19}. Fig. \ref{fig:Model-vs-Obs} shows the model dust and gas surface density profiles from fig. \ref{fig:TD-3Wavelengths} that we found to match the observed disc spectra best. These are compared with the broken power-law dust surface density model of \cite{Hogerheijde16-TWHYA} and the three models for $\Sigma$ previously shown in fig. \ref{fig:Obs}.

We see that while the dust surface density match is reasonably good (which of course is the goal of our paper), the gas surface density profile and the total gas disc mass is significantly lower than previous workers assumed or derived. Without sophisticated modelling it is unclear how serious the disagreement is. Since the disrupted planet is metal rich in our scenario, the mass of various molecular species may be sufficiently high in the model to account for their observed emission since previous workers assumed much lower abundances for the disc. However, HD line emission is not expected to be sensitive to the metallicity of the gas \citep[e.g.,][]{BerginEtal13,WoitkeEtal19}; it remains to be seen whether our much less massive disc may account for the observed line fluxes. 

\section{Conclusions}

Here we focused on the first ALMA 1.3 mm dust continuum {\em excess} emission \citep{TsukagoshiEtal19} positioned right at the edge of a cliff-like rollover of the dust disc in TW Hydra. We showed that the morphology of the blob-like excess and its relation to the dust disc are best explained by a planet losing dust and gas within the excess. We argued that pre-runaway Core Accretion planets and pre-collapse Gravitational Instability planets may be disrupted and may provide the required mass injection into the system. This catastrophic event may also explain why there is a factor of $\sim 100$ more dust in this very old system than the mean for (typically younger) class II protoplanetary discs. Future modelling needs to improve on the internal planet structure, mass loss dynamics, dust composition and opacity, and chemodynamical modelling.

\section*{Acknowledgements}
The authors thank John Ilee for useful discussions and suggestions. SN acknowledges support from STFC grants ST/N000757/1 and ST/M006948/1 to the University of Leicester. This work made use of the DiRAC Data Intensive service at Leicester, operated by the University of Leicester IT Services, which forms part of the STFC DiRAC HPC Facility (www.dirac.ac.uk). The equipment was funded by BEIS capital funding via STFC capital grants ST/K000373/1 and ST/R002363/1 and STFC DiRAC Operations grant ST/R001014/1. DiRAC is part of the National e-Infrastructure. CH is a Winton Fellow and this work has been supported by Winton Philanthropies, The David and Claudia Harding Foundation. FM acknowledges support from the Royal Society Dorothy Hodgkin Fellowship.



\bibliographystyle{mnras}
\bibliography{nayakshin}


\appendix

\section{Dust opacity uncertainties}\label{sec:App-dust-opacities}

\begin{figure}
\includegraphics[width=0.48\textwidth]{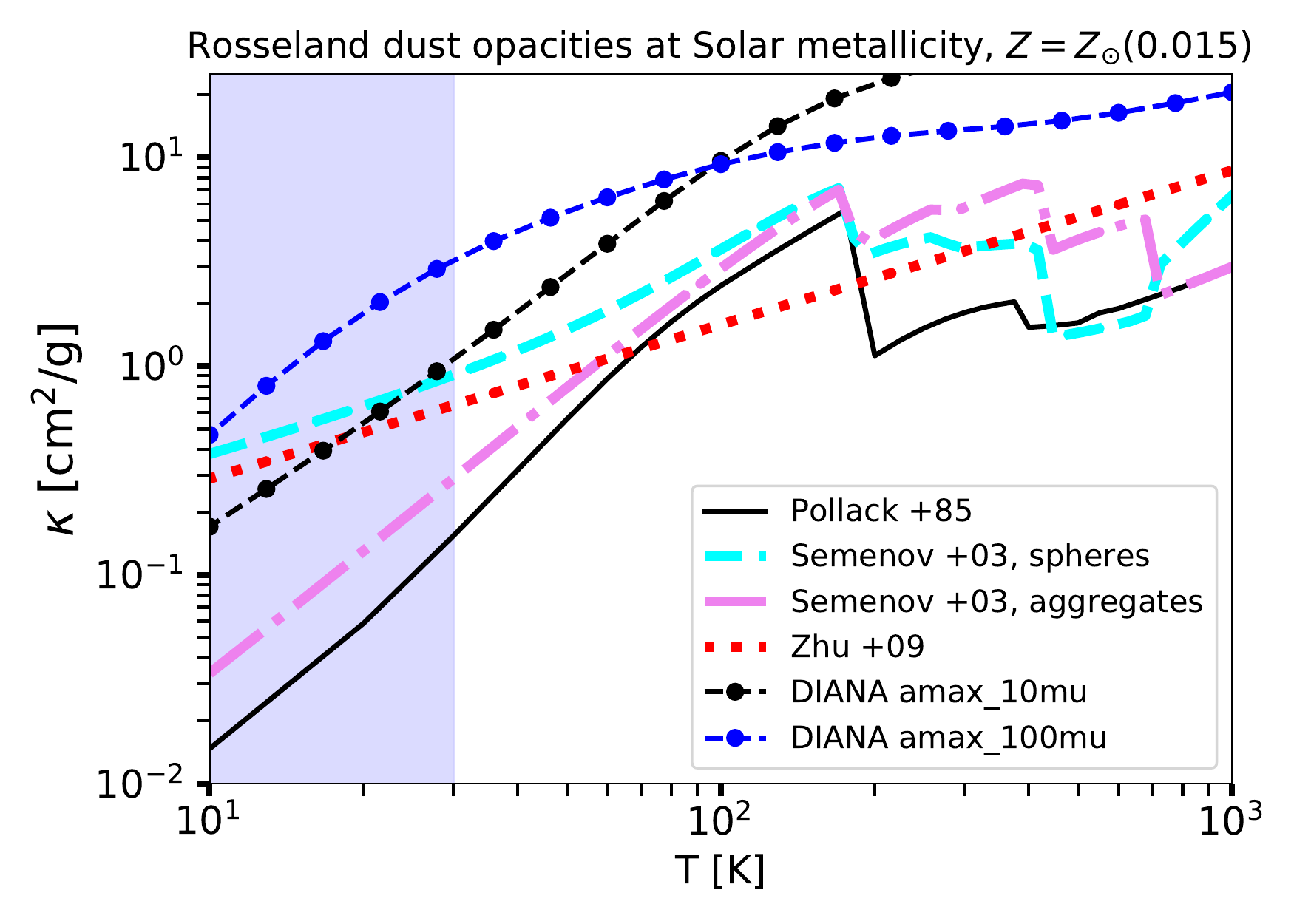}
\caption{Rosseland opacity as a function of temperature for different dust opacity models as indicated in the legend.}
\label{fig:Compare_Kappas}
\end{figure}

Fig. \ref{fig:Compare_Kappas} shows several models for the absorption Rosseland mean opacity, $\kappa_{\rm R}$, of gas at Solar metallicity from several different authors as a function of gas temperature. For temperatures $T\le 10^3$~K and gas densities typical of pre-collapse planets, $\kappa_{\rm R}$ is strongly dominated by dust. For the present paper, the shaded temperature region, $10 \le T \le 30$~K, is most important as this encompasses the expected effective temperatures for our dust-rich wide-separation planets \citep[the energy transfer in deeper hotter planet interiors is dominated by convection anyway][]{HelledEtal08}. The two DIANA opacity calculations \citep{Woitke16-DIANA} neglect grain vaporisation as this is not important in the shaded region, but include the effects of grain growth, by allowing the maximum grain size to be either $10 \mu$m or $100 \mu$m.

We see that there is a factor of about 30 uncertainty between the smallest and the largest $\kappa_{\rm R}$. This shows that early calculations of giant planet contraction \citep[e.g.,][]{Bodenheimer74} may have significantly over-estimated the luminosity of these objects. Furthermore, higher metallicity objects have proportionally higher dust opacities, further delaying planet contraction. 

\section{Pre-disruption planet contraction computed with two different codes}\label{sec:App-planet-disruption}

\begin{figure}
\includegraphics[width=0.48\textwidth]{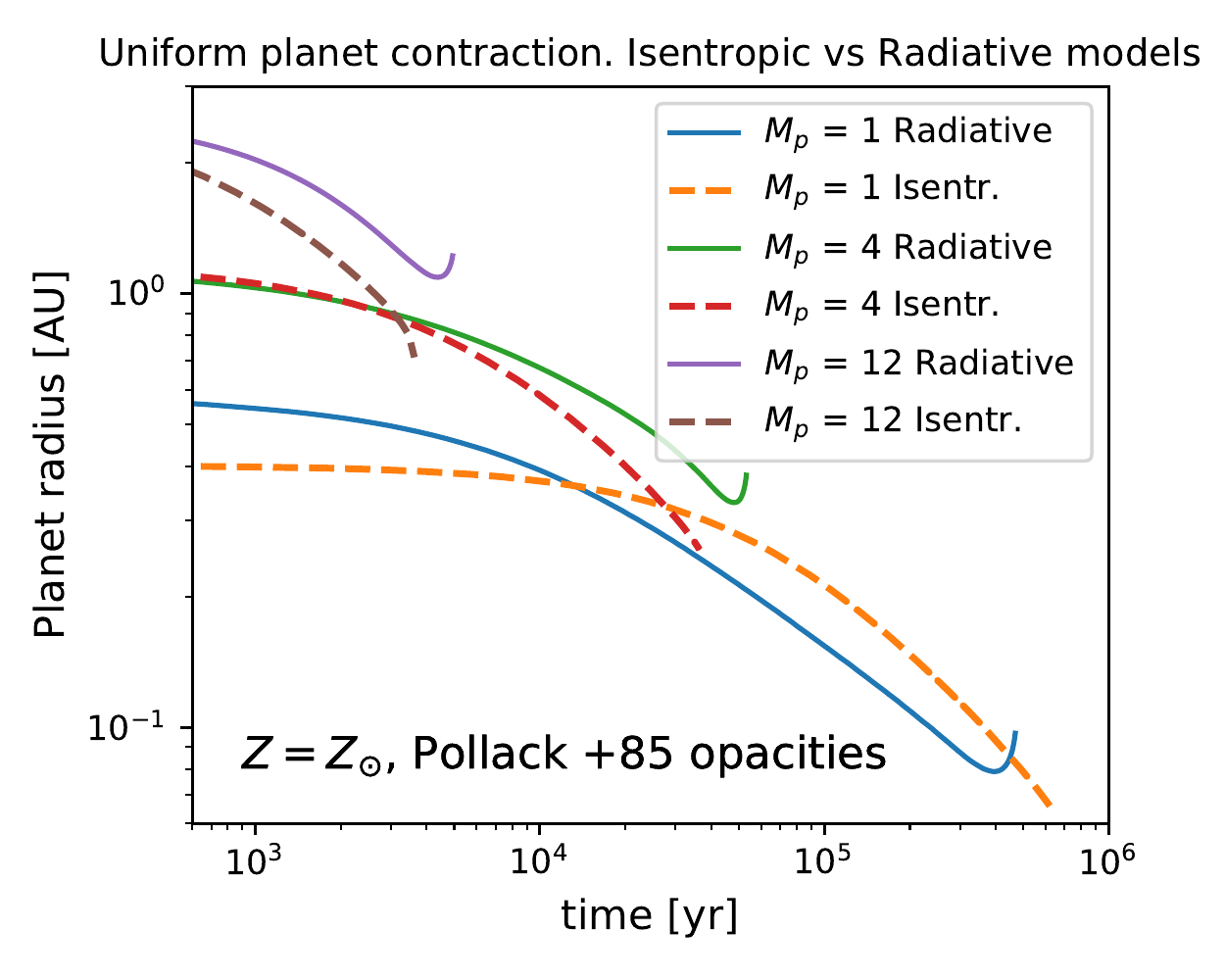}
\caption{Comparison of planet radii versus time computed with the codes of Nayakshin (2016) and Vazan \& Helled (2012). For each planet mass, the two models are initialised with the same central temperature.}
\label{fig:Compare_Allona_vs_mine}
\end{figure}

As explained in \S \ref{sec:GI-core-disruption}, 
to model planet contraction simultaneously with dust growth and sedimentation into the core, we use the code of \cite{Nayakshin16a}. Here we compare the results of this code, which uses an isentropic (follow-adiabats) approximation to the energy transfer through the planet envelope, to the more accurate stellar evolution model of \cite{VazanHelled12} for the simpler case in which grain growth and sedimentation are neglected. Fig. \ref{fig:Compare_Allona_vs_mine} shows the evolution of planetary radius computed with the two different codes for the same opacity \citep{Pollack85-Opacity} for several planet masses. The evolutionary tracks computed with the two codes are within $\sim 30$\% of each other in terms of the absolute value of the planet radius, and within a factor of two in terms of the planet collapse time scales. We deem this sufficiently close given the much larger uncertainty that exists in the dust opacity. 

\bsp	
\label{lastpage}
\end{document}